\documentclass[11pt,a4paper,epsf,epsfig,psfrag]{article}
\usepackage{jheppub}
\usepackage{amsmath,graphicx,amsfonts, mathrsfs,amssymb}
\usepackage{amsmath,epsfig}
\usepackage{amssymb,amsfonts}
\usepackage{color}
\usepackage{latexsym}
\usepackage{epsfig}
\usepackage{pdfsync}
\usepackage{marvosym}
\newbox\pippobox

\title{Quarkyonic phase from quenched dynamical holographic QCD model}
\author[a,b,1]{Xun Chen,\note{first author,}}
\author[c,2]{Danning Li,\note{co-corresponding author,}}
\author[b]{Defu Hou, }
\author[a,3]{and Mei Huang \note{corresponding author.}}

\affiliation[a]{School of Nuclear Science and Technology, University of Chinese Academy of Sciences,\\Beijing 100049, P.R. China}
\affiliation[b]{Central China Normal University, Wuhan 430079, P.R. China }
\affiliation[c]{Department of Physics and Siyuan Laboratory, Jinan University, Guangzhou 510632, P.R. China}
\emailAdd{chenxunhep@qq.com}
\emailAdd{lidanning@jnu.edu.cn}
\emailAdd{houdf@mail.ccnu.edu.cn}
\emailAdd{huangmei@ucas.ac.cn}

\abstract{Chiral and deconfinement phase transitions at finite temperature $T$ and quark number chemical potential $\mu$ are simultaneously studied in the quenched dynamical holographic QCD model within the Einstein-Dilaton-Maxwell framework. By calculating the corresponding order parameters, i.e., the chiral condensate and Polyakov loop,  it is shown that the transition lines of these two phase transitions are separated in the $T-\mu $ plane. The deconfinement phase transition is shown to be always of crossover type and the transition line depends weakly on the baryon number density. Differently, the chiral transition is of crossover at small baryon number density and it turns to be of first order at sufficient large baryon number density.  A critical endpoint (CEP), at which the transition becomes second order type, appears in the chiral transition line. This is the first time to realize the CEP of chiral phase transition in the $(T, \mu)$ plane using the holographic EMD(Einstein-Maxwell-Dilaton) model for two flavour case. It is observed that between these two phase transition lines, there is a region with chiral symmetry restored and color degrees still confined, which could be considered as the quarkyonic phase. Qualitatively, this behavior is in consistent with the result in the Polyakov-loop improved Nambu-Jona-Lasinio (PNJL) model.}


\begin{document}
\maketitle
\section{Introduction}
\label{sec-int}

Quantum Chromodynamics (QCD) is widely accepted as the fundamental theory of the strong interaction with two most important properties  in the vacuum, i.e., the spontaneous chiral symmetry breaking and color confinement. At sufficient high temperature or/and baryon chemical potential, it is believed that the system will undergo phase transitions, involving the restoration of chiral symmetry and the release of color degrees of freedom. The interplay of chiral symmetry breaking and color confinement as well as the relation between chiral and deconfinement phase transitions at finite temperature and density reveal the fundamental property of quark dynamics and gluon dynamics, thus it attracts continuous interests \cite{Polyakov:1978vu,'tHooft:1977hy,Casher:1979vw,Banks:1979yr,Hatta:2003ga,Mocsy:2003qw,Braun:2007bx}.
In the limit of large number of colors $N_c$,  the quarkyonic phase was expected \cite{McLerran:2007qj,Hidaka:2008yy,McLerran:2008ua} in certain baryon number density region, where the chiral symmetry is restored but color degrees of freedom are still confined.

It requires more efforts to understand the full properties of QCD, since chiral symmetry breaking and color confinement have a non-perturbative origin, when the traditional perturbative methods face enormous challenges. Lattice QCD, starting from the first principle at zero and small quark chemical potential \cite{Aoki:2006br,Schmidt:2006us,Philipsen:2005mj,Heller:2006ub}, is regarded as an important tool to overcome the non-perturbative problem. Besides lattice QCD, other non-perturbative methods such as Dyson-Schwinger equations (DSEs) \cite{Alkofer:2000wg,Bashir:2012fs}, functional renormalization group equations (FRGs) \cite{Wetterich:1992yh,Pawlowski:2005xe,Gies:2006wv} and QCD effective models have been developed. Among QCD low energy effective models, Nambu--Jona-Lasinio (NJL) model \cite{Nambu:1961fr,Klevansky:1992qe} offers the mechanism of spontaneous chiral symmetry breaking and has been widely used in describing chiral phase transition and investigating QCD phase structures under variant extreme conditions. In this model, the QCD gluon-mediated interactions are replaced by effective interactions among quarks, which are built according to the global symmetries of QCD. The NJL model does not contain dynamical gluons, which can be improved by adding the Ginzburg-Landau type potential of the traced Polyakov loop to the lagrangian to describe gluon dynamics and an interaction term of the Polyakov loop with the quarks. The improved model is usually named the Polyakov-loop improved Nambu--Jona-Lasinio(PNJL) model \cite{Meisinger:1995ih,Fukushima:2003fw,Ratti:2005jh,Roessner:2006xn,Ghosh:2006qh,Schaefer:2007pw,Ratti:2007jf,Sasaki:2006ww,Sasaki:2006ws,Megias:2006bn,Megias:2004hj,Zhang:2006gu,Sakai:2008py,Ciminale:2007ei,Fu:2007xc,Hansen:2006ee,Contrera:2007wu}.

In recent decades, the discovery of the anti-de Sitter/conformal field theory (AdS/CFT) correspondence and the conjecture of the gauge/gravity duality \cite{Maldacena:1997re,Gubser:1998bc,Witten1998Anti} leads a new way to solve the strong coupling problem of gauge theory. Comparing with the original AdS/CFT correspondence, it is necessary to break the conformal symmetry at low energy to describe QCD. Many efforts have been made towards more realistic holographic description of low energy phenomena of QCD in hadron physics \cite{Erlich:2005qh,deTeramond:2005su,DaRold:2005mxj,Babington:2003vm,Kruczenski:2003uq,Sakai:2004cn,Sakai:2005yt,Csaki:2006ji,Huang:2007fv,Gherghetta:2009ac,Kelley:2010mu,Sui:2009xe,Sui:2010ay,Li:2012ay,Li:2013oda,Chen:2015zhh} and hot/dense QCD \cite{Shuryak:2004cy,Tannenbaum:2006ch,Policastro:2001yc,Cai:2009zv,Cai:2008ph,Sin:2004yx,Shuryak:2005ia,Janik:2005zt,Nakamura:2006ih,Sin:2006pv,Herzog:2006gh,Gubser:2006bz,Zhang:2012jd,Li:2014dsa,Li:2014hja}, including the top-down approaches and bottom-up approaches (see \cite{Aharony:1999ti,Erdmenger:2007cm,Brodsky:2014yha,Kim:2012ey,Adams:2012th}for reviews). The deconfinement phase transition \cite{Herzog:2006ra,BallonBayona:2007vp,Cai:2007zw,Cai:2012eh,Kim:2007em,Andreev:2009zk,Colangelo:2010pe,Gubser:2008yx,Gubser:2008ny,Gubser:2008sz,Gursoy:2008bu,Gursoy:2007cb,Gursoy:2007er,Gursoy:2008za,Yaresko:2013tia,Li:2011hp,Cai:2012xh,He:2013qq,Yang:2014bqa,Zuo:2014iza,Afonin:2014jha,Bruckmann:2013oba,Rougemont:2015oea}
has been widely discussed with the expectation value of Polyakov loop is the order parameter of deconfinement phase transition. However, till recently, the dynamical spontaneous chiral symmetry breaking and chiral phase transition have been realized \cite{Karch:2006pv,Chelabi:2015cwn,Chelabi:2015gpc,Fang:2015ytf,Evans:2016jzo,Evans:2010iy,Evans:2012cx,Evans:2011eu} with the chiral condensate as the order parameter.

 In large $N_c$ limit, it has been proposed that the chiral symmetry and deconfinement phase transition can be splitting and there would be a quarkyonic phase \cite{McLerran:2007qj},  where the chiral symmetry is restored but still in confinement. The free energy of a heavy test quark added to the system is $e^{-\beta F_{q}} = \frac{1}{N_{c}} <L>$ and the baryon number density is $<N_B> \sim e^{-\beta(M_B-\mu_B)}$ with $\beta$ the inverse of temperature. The deconfinement temperature is around $T_d \sim \Lambda_{QCD}$.When the baryon number chemical  potential is small compared to the baryon mass, i.e. $\mu_B \ll M_B $ , the number of baryons $<N_B> \sim e^{- \kappa_1 N_c}$ tends to zero at large $N_c$ for temperatures of $T \sim \Lambda_{QCD}$ and with $M_{B} \sim N_c$. Here, $\kappa_1$ is a number of order one. When $\mu_B \geq M_B$, the baryons begin to populate the system and the baryon number density becomes nonvanishing. Ref.\cite{McLerran:2007qj} argues that there are at least three phases in the QCD phase diagram at large $N_c$: the deconfinement phase with $T>T_d$, and in the region of $T<T_d$, there will be the mesonic phase which is confined and has vanishing baryon number density and the quarkyonic phase which has finite baryon number density and is confined \cite{McLerran:2008ua}. However, in the real QCD case with $N_c=3$, we normally call the chiral symmetric and confined phase as the quarkyonic phase as shown in the PNJL model  \cite{McLerran:2008ua,Sakai:2012ika}. 

 In this paper, we make a step towards investigating the interplay between the deconfinement phase transition and chiral phase transition in a quenched dynamical holographic QCD model \cite{Li:2012ay,Li:2013oda,Li:2014dsa}. In this quenched dynamical holographic QCD model, the dilaton background describes the gluodynamics and the flavor/meson background describes the chiral dynamics, respectively, thus one can simultaneously realize the confinement/deconfinement phase transition and chiral symmetry breaking/restoration phase transition at finite temperature and chemical potential. However, it is worthy of mentioning that in the quenched dynamical holographic QCD model, the flavor background is added on the dilaton background as a probe, and the full QCD dynamics including the backreaction from the flavor background on the dilaton background or gluodynamics background has not been self-consistently solved yet, which is left for future work.

To extend the quenched dynamical holographic QCD model to finite chemical potential, the quark chemical potential is introduced by a $U(1)$ field in the Einstein-Dilaton-Maxwell framework. Except to fix the chemical potential dependence of the flavor background, one has also to fix the chemical potential dependence of the dilaton/gluodynamics potential, which can be determined by higher order baryon number fluctuations especially the kurtosis of baryon number fluctuations. From the experience in the PNJL model \cite{Li:2018ygx} as well as in the holographic QCD model \cite{Li:2017ple},  the kurtosis of baryon number fluctuations is dominated by contribution from gluodynamics. Therefore, we fix the chemical potential dependence of the dilaton field by fitting the lattice results of the kurtosis of baryon number fluctuations \cite{Bazavov2017The} at zero chemical potential. It is found that the deconfinement phase transition in the $(T,\mu)$ plane is always a crossover, this is in agreement with the result in the PNJL model \cite{Li:2018ygx}. By adding the coupling of $\chi^6$ with chemical potential, the critical end point (CEP) shows up along the chiral phase transition line. It is also found that the chiral symmetry restoration and deconfinement phase transitions are separated. More interestedly, it is observed that there exists a region where chiral symmetry is restored but color degrees of freedom are still confined. This is similar to the quarkyonic phase obtained in the PNJL model \cite{Li:2018ygx}. The possible reason for the separation of the chiral and deconfinement phase transitions is due to the quenched gluodynamical background, where the flavor dynamics is added as a probe.

The remainder of the paper is organized as follows: We give a review on the quenched dynamical holographic QCD model in Sec.\ref{sec-action}. In Sec.\ref{deconfinement}, we fix the chemical potential dependence of the dilaton potential which describes gluodynamics through the baryon number susceptibilities and investigate the deconfinement phase transition in the $(T,\mu)$ plane. In Sec.\ref{chiral} we investigate the chiral phase transition and the phase diagram in the quenched dynamical holographic QCD model. In Sec.\ref{sec-sum}, a brief summary is given.

\section{Setup for the quenched dynamical holographic QCD model at finite baryon chemical potential}
\label{sec-action}

The deconfinement phase transition has been widely investigated in bottom-up holographic models \cite{Herzog:2006ra,BallonBayona:2007vp,Cai:2007zw,Cai:2012eh,Kim:2007em,Andreev:2009zk,Colangelo:2010pe,Gubser:2008yx,Gubser:2008ny,Gubser:2008sz,Gursoy:2008bu,Gursoy:2007cb,Gursoy:2007er,Gursoy:2008za,Yaresko:2013tia,Li:2011hp,Cai:2012xh,He:2013qq,Yang:2014bqa,Zuo:2014iza,Afonin:2014jha}. Besides, it is possible to describe the chiral symmetry breaking and its restoration in the soft-wall model
\cite{Karch:2006pv,Chelabi:2015cwn,Chelabi:2015gpc}. In this work, we investigate the interplay between the chiral and deconfinement phase transitions by using the dynamical holographic QCD model \cite{Li:2012ay,Li:2013oda,Chen:2015zhh,Li:2014dsa,Li:2013xpa,Li:2014txa}.  We emphasize that though the original AdS/CFT correspondence is in the large $N_c$ limit, the holographic QCD model is trying to describes phenomenology of real QCD in the case of $N_c=3$. In some sense, this is under a more general holographic conjecture for gauge/gravity duality \cite{tHooft:1993dmi,Susskind:1994vu}. 

The full QCD contains quark dynamics and gluodynamics, and it is known that light flavor quark dynamics are responsible for the spontaneous chiral symmetry breaking, and gluodynamics are responsible for the color confinement. The dynamical holographic QCD model is constructed in the graviton-dilaton-scalar framework with the dilaton field and scalar field responsible for the gluodynamics and chiral dynamics, respectively. This dynamical holographic QCD model naturally resembles the renormalization group from ultraviolet (UV) to infrared (IR): at UV boundary, the theory goes to the limit of AdS/CFT, and the 5-dimension (5D) field in the bulk and 4D operator obeys the principle of AdS/CFT correspondence \cite{Erlich:2005qh}, and the model at IR is determined by QCD non-perturbative properties such as chiral condensate and gluon condensate or glueball properties. In \cite{Li:2012ay,Li:2013oda,Chen:2015zhh,Li:2014dsa,Li:2014hja}, we see that the graviton-dilaton system can describe the pure gluon system including the glueball spectra, thermodynamical properties as well as transport properties quite well. After adding the flavor background and solving the deformed metric self-consistently, the total dynamical system can describe the meson spectra very well and the results are in agreement with experimental data \cite{Li:2013oda}.

However, it is not an easy task to solve the full system at finite temperature and chemical potential. Therefore, in this work, we use the quenched dynamical model with the flavor background added on the dilaton background as a probe. The chiral and deconfinement phase transitions at finite temperature in the quenched dynamical holographic QCD model has been investigated in Ref. \cite{Fang:2015ytf}. Here, we extend this scenario to finite chemical potential case, and try to study chiral and deconfinement phase transition in the $T - \mu$ plane.

 To extend the quenched dynamical holographic QCD model in Refs. \cite{Li:2012ay,Li:2013oda,Li:2014dsa} to finite chemical potential , we introduce an extra $U(1)$ field in the Einstein-Dilaton-Maxwell framework, and the action in the string frame takes the form of:
 \begin{eqnarray}
 S_{total}^s&&=S_G^s+S_M^s,\\
S_G^s &&= \frac{1}{16\pi G_5}\int d^5x\sqrt{-g^s}e^{-2\Phi}[R^s+4\partial_\mu\Phi\partial^\mu\Phi-V_G^s(\Phi)-\frac{h(\Phi)}{4}e^\frac{4\Phi}{3} F_{\mu\nu}F^{\mu\nu}]\label{sb},\\
S_M^s &&= -\int d^5x\sqrt{-g^s}e^{-\Phi}Tr[\nabla_\mu X^{\dagger}\nabla^\mu X+V_X^s(|X|,F_{\mu\nu}F^{\mu\nu} )]. \label{SMatter}
\end{eqnarray}
Here $S_{total}$ is the full 5D action, $S_G$ is the 5D action for dilaton background describing gluodynamics, and $S_M$  is the 5D action for matter sector describing chiral dynamics, respectively. The lower-case $s$ represents the string frame, $g^s$ is the determinant of metric $g_{\mu\nu}$, $G_5$ is the 5D Newton constant, $\Phi$ is the dilaton field, and $X$ is the bulk scalar field which corresponds to $\bar{q}q$ condensate of QCD. $V_G$ represents the dilaton potential, and $V_X$ is the bulk scalar potential coupled with the strength tensor of gauge field. The leading term in $V_X$ is the mass term $m_{5}^{2}XX^{\dagger}$, which can be determined as $m_{5}^{2} = -3$ from the AdS/CFT prescription $m_{5}^{2} = (\Delta - p)(\Delta + p - 4)$ by taking $\Delta = 3, p = 0$ \cite{Witten1998Anti,BallonBayona:2007vp}. $h(\Phi)$ is a gauge kinetic function constraining the $\mu$ dependence of the system and will be fixed by fitting the lattice data on baryon number susceptibilities. $F_{\mu\nu}$ are the strength tensor of gauge field dual to the baryon number current. If $F_{\mu\nu}=0$,  the system is reduced to zero chemical potential case, and $F_{\mu\nu}\neq0$ corresponds to finite baryon number chemical potential case of the system.

To consider the gravity dual of QCD at finite temperature and baryon number density, we can take the following metric ansatz in the string frame:
\begin{eqnarray}
ds^2=\frac{e^{2A_s(z)}}{z^2}[-f(z)dt^2+\frac{1}{f(z)}dz^2+d\vec{x}^2]. \label{metrics}
\end{eqnarray}
As discussed in Ref.\cite{Li:2011hp}, it is more convenient to work out thermodynamics in the Einstein frame, therefore we transform the action into Einstein frame by a conformal transformation of metric:
\begin{eqnarray}
g^{s}_{\mu\nu}=e^{\frac{4\Phi}{3}}g^{e}_{\mu\nu},
\end{eqnarray}
then the action of the dilaton background part becomes:
\begin{eqnarray}
S_G^e &&= \frac{1}{16\pi G_5}\int d^5x\sqrt{-g^e}[R^e-\frac{4}{3}\partial_\mu\Phi\partial^\mu\Phi-V^e(\Phi)- \frac{h(\Phi)}{4}F_{\mu\nu}F^{\mu\nu}],
\label{S_G_E}
\end{eqnarray}
with
\begin{equation}
V^e(\Phi)=e^{4/3\Phi}V^s(\Phi).
\end{equation}
In the Einstein frame, the metric ansatz becomes:
\begin{eqnarray}
ds^2=\frac{e^{2A_e(z)}}{z^2}[-f(z)dt^2+\frac{1}{f(z)}dz^2+d\vec{x}^2].\label{metrice}
\end{eqnarray}
Here, the two metric warp factors in two frames follow the relationship of $A_s = A_e + \frac{2\Phi}{3}$. When considering finite chemical potential, the only non-vanishing component of the gauge potential $A_\mu$  of the strength tensor $F_{\mu\nu}$ is the time component $A_t$, i.e. $A=A_t dt$.

Inserting the above ansatz, one can derive the following equations of motion after certain simplifications \cite{Fang:2015ytf}:
\begin{eqnarray}
A^{\prime\prime}_e-A^{\prime2}_e+\dfrac{2}{z}A^{\prime}_e+\dfrac{4\Phi^{\prime2}}{9}
&  =0,\label{eom-A}\\
A_{t}^{\prime\prime}+\left(  \frac{h^{\prime}}{h}+A^{\prime}_e-\dfrac{1}%
{z}\right)  A_{t}^{\prime}  &  =0,\label{eom-At}\\
f^{\prime\prime}+\left(  3A^{\prime}_e-\dfrac{3}{z}\right)  f^{\prime}%
-e^{-2A_e}z^{2}h(\Phi)A_{t}^{\prime2}  &  =0.\label{eom-f}
\end{eqnarray}
The Hawking temperature of the black hole solution can be evaluated numerically by
\begin{eqnarray}
T = \frac{f'(z_h)}{4\pi}.
\end{eqnarray}

Following \cite{Li:2013oda},  we take the dilaton field in the form of
\begin{equation}
\Phi(z) =\alpha\ {\rm tanh} (\beta^2z^2+\gamma^4z^4),
 \end{equation}
which tends to $z^2$ power thus goes to the AdS$_5$ limit at the UV boundary and approaches a positive constant for a possible crossover transition at IR. Then one can solve Eqs.(\ref{eom-A})-(\ref{eom-f}) by imposing the following boundary conditions at the boundary $z=0$ and the horizon $z = z_h$:
\begin{eqnarray}
A_t(z_h) &=& f(z_h)=0,\\
f(0) &=& 1,\\
A_t(0) &=& \mu + \rho z^2 +....
\end{eqnarray}
Here, $\mu$ is the quark chemical potential and $\rho$ is quark number density.

Following the procedure in Ref.\cite{Li:2011hp}, we can calculate the pressure of the system through the entropy density
\begin{eqnarray}
s = \frac{e^{3A(z_h)}}{4z^3_{h}}.
\end{eqnarray}
For fixed values of the chemical potential, the pressure density can be calculated by the integral
\begin{eqnarray}
p = \int [s dT+\rho d\mu]=\int[s (\frac{\partial T}{\partial z_h}d z_h+\frac{\partial T}{\partial \mu}d \mu)+\rho d\mu]
\end{eqnarray}
and the energy density of the system can be derived:
\begin{equation}
\epsilon = -p + sT + \mu \rho.
\end{equation}

At zero chemical potential $\mu=0$, by fitting the lattice results of equation of state for $N_f = 2$ QCD\cite{Burger2015The}, one can fix parameters $\alpha = 1.8, \beta = 0.4\ {\rm GeV}, \gamma = 0.42\ {\rm GeV}$ \textcolor{blue}{as in Ref.\cite{Fang:2015ytf}}. With these parameters, the pseudo critical temperature for the crossover is around $T_0=217~ {\rm MeV}$ \cite{Fang:2015ytf}.

In order to describe the system at finite chemical potential, we also need to fix $h(\Phi)$ in Eq.(\ref{sb}) and (\ref{S_G_E}), which describes the chemical potential dependence of the gluon-dynamical potential of the system. From the experience in the PNJL model \cite{Li:2018ygx}, we can use the higher order baryon number fluctuations especially the ratio of fourth over second order cumulants of baryon number fluctuations to determine $h(\Phi)$. The kurtosis of baryon number fluctuations is given by $\kappa\sigma^2=C_4^B/C_2^B$ with the variance $\sigma^2=C_2^B$ and the kurtosis $\kappa={C_4^B}/{(\sigma^2)^2}$, and the cumulants of baryon number distributions are given by $C_n^B=VT^3\chi^B_n$, where the baryon number susceptibilities are defined as:
\begin{eqnarray}
\chi^B_n = \frac{\partial^n[P / T^4]}{\partial[\mu_B / T]^n},
\end{eqnarray}
and $P,V$ are the pressure and volume of the system, and $\mu_B=3 \mu$ is the baryon number chemical potential. The calculation of baryon number fluctuations in the holographic QCD model of \cite{Yang:2014bqa} has been investigated in Ref. \cite{Li:2017ple}.
The baryon number fluctuations in the PNJL model has been investigated in Ref. \cite{Li:2018ygx} and it is observed that in the PNJL model \cite{Li:2018ygx} that the ratio of fourth over second order cumulants of net-baryon number fluctuations $\kappa\sigma^2$ at zero baryon number density is dominated by contribution from gluodynamics. Therefore, we fix the chemical potential dependence of the dilaton field by fitting the lattice results of the kurtosis of baryon number fluctuations \cite{Bazavov2017The} at zero chemical potential.

\begin{figure}
\centering
    \includegraphics[width=6.5cm,height=4.5cm]{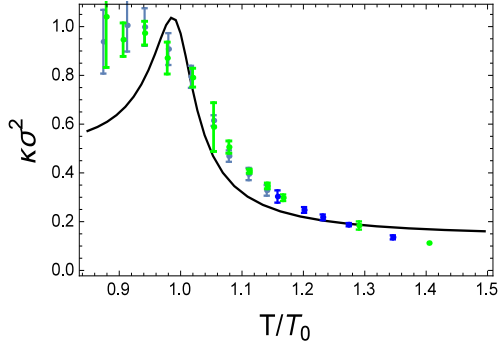}
\caption{$\kappa\sigma^2$ as functions of the normalized temperature $T/T_0$ with $T_0=217~{\rm MeV}$ the pseudo critical temperature at zero chemical potential $\mu=0$. The black line is for $h(\Phi)=7/10$ which is in agreement with lattice results in Ref.\cite{Bazavov2017The}.
 }
 \label{chi}
\end{figure}

The kurtosis $\kappa\sigma^2$ as a function of the normalized temperature $T/T_0$ with $T_0=217~{\rm MeV}$ comparing with lattice result \cite{Bazavov2017The} is shown in Fig.\ref{chi}. It is found that when $h(\Phi) = \frac{7}{10}$, the result of kurtosis from the dilaton background, i.e., from the gluodynamical contribution in the quenched dynamical holographic QCD model is in good agreement with lattice results.

With the setup in this section, we are ready to investigate the deconfinement and chiral phase transitions in the quenched dynamical holographic QCD model at finite temperature as well as finite chemical potential.

\section{Deconfinement transition at finite baryon density}
\label{deconfinement}

In order to investigate the deconfinement phase transition, the expectation value of the Polyakov loop is often used as an order parameter, which is defined as
\begin{eqnarray}
L(T) = \frac{1}{N_c}{\rm Tr}{\rm P}~ exp(ig\int_{0}^{\frac{1}{T}}\hat{A_0}(\tau,\vec{x})\,d\tau).
\end{eqnarray}
Here $N_c$ is the color number, P indicates path ordering, g is the coupling, the trace ${\rm Tr}$ is computed over the fundamental representation of $SU(N_c)$ and $\hat{A_0}$ is the non-Abelian gauge field potential operator. The expectation value $\langle L(T)\rangle$ vanishes in the confined phase guaranteed by the center symmetry, and it is nonzero $\langle L(T)\rangle  \neq 0$ in the deconfined phase, which indicates the center symmetry is broken. From the holographic dictionary, the expectation value of the Polyakov loop is related to  the Nambu-Goto action $S_{NG}$ for the string world sheet
\cite{Witten1998Anti}
\begin{equation}
S_{NG} = \frac{1}{2\pi\alpha_p}\int d^2\eta\sqrt{{\rm det}(g^s_{\mu\nu}\mathcal{X}^\mu_a\mathcal{X}^\nu_b)}
\end{equation}
in the following way
\begin{equation}
\langle L(T)\rangle = \int D \mathcal{X} e^{-S_{NG}},
\end{equation}
where $\frac{1}{2\pi\alpha_p}$ is the string tension, $g^s_{\mu\nu}$ is the metric in the string frame, and $\mathcal{X}^\mu_a$ is the embedding function of the worldsheet in the target space, $\mu,\nu$ are the five dimensional space-time indices and $a,b$ represent the worldsheet coordinates. From the metric of \eqref{metrics}, we have
\begin{eqnarray}
S_{NG} = \frac{1}{2\pi\alpha_p}\int_{0}^{z_h} dz \frac{e^{2A_s}}{z^2}\sqrt{1+f(z)(\vec{x}^\prime)^2},
\end{eqnarray}
with $g_p = \frac{1}{2\alpha_p}$ the redefinition of the string tension. The prime denotes the derivative with respect to $z$. Then, the equation of motion for $\vec{x}$ can be derived as
\begin{eqnarray}
[{\frac{e^{2A_s}}{z^2}f(z)\vec{x}^\prime/\sqrt{1+f(z)(\vec{x}^\prime)^2}}]^\prime = 0.
\end{eqnarray}

Substituting the $\vec{x}^\prime$ into the action $S_{NG}$, the minimal world sheet can be obtained as
\begin{eqnarray}
S_0 = c_p +S^{\prime}_0 = c_p +\frac{g_p}{\pi T}\int^{z_h}_{0} dz(\frac{e^{2A_s}}{z^2}-\frac{1}{z^2}),
\end{eqnarray}
where $c_p$ is a normalization constant. Finally, we can get the expectation value of the Polyakov loop:
\begin{eqnarray}
\langle L(T)\rangle = \omega e^{-S_0} = e^{C_p - S^\prime_0},
\end{eqnarray}
with $C_p$ another normalization constant and $\omega$ a weight coefficient.
By fitting two-flavor lattice results from \cite{Burger2013The}, we take $C_p = -0.25, g_p=0.86.$

\begin{figure}
\begin{center}
\epsfxsize=6.5 cm \epsfysize=4.5 cm \epsfbox{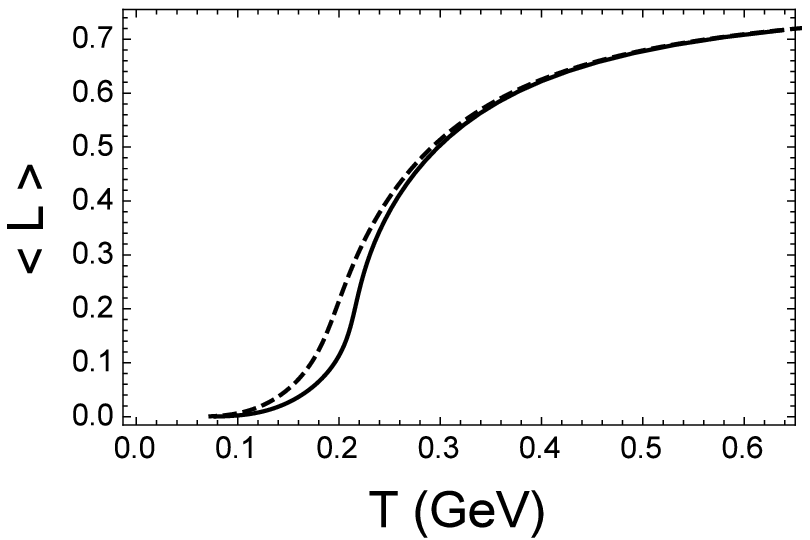}
\hspace*{0.1cm} \epsfxsize=6.5 cm \epsfysize=4.5 cm
\epsfbox{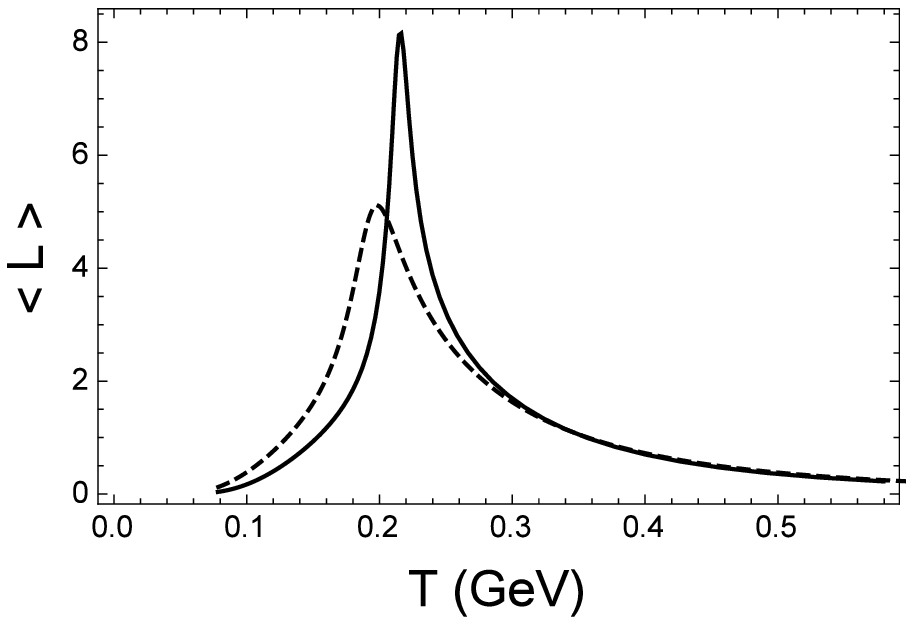} \vskip -0.05cm \hskip 0.15 cm
\textbf{( a ) } \hskip 6.5 cm \textbf{( b )} \\
\end{center}
\caption{The expectation value of the Polyakov loop $< L >$ (a) and its derivative $\frac{d< L >}{dT}$ (b) as a function of the temperature in the cases of $\mu = 0$ (solid line) and $\mu = 0.5\ {\rm GeV}$(dashed line), respectively.}
 \label{L_mu0_mu05}
\end{figure}

\begin{figure}
\centering
\includegraphics[width=6.5cm,height=4.5cm]{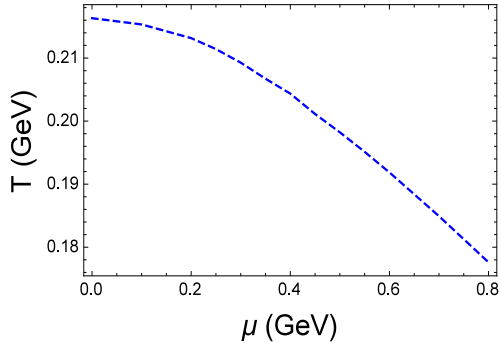}
\caption{The deconfinement phase transition line in the $(T,\mu)$ plane in the quenched dynamical holographic QCD model.}
\label{Tmu1}
\end{figure}

The expectation value for the Polyakov loop $\langle L\rangle$ at the chemical potentials $\mu = 0$ and $\mu = 0.5  {\rm GeV}$ are given in Fig.\ref{L_mu0_mu05}(a). We can see that at low temperature the system is in confined phase with large $\langle L \rangle$, while at large temperature it tends to zero showing a deconfinement phase transition. The transition from the confined phase to the deconfined phase is smooth, showing a crossover type transition.  Usually, one can extract the corresponding pseudo critical temperatures through  the derivative of $\frac{d \langle L\rangle}{dT}$, and the location of the peak gives the pseudo critical temperature. In Fig.\ref{L_mu0_mu05}(b),  the results for $\frac{d \langle L\rangle}{dT}$ are given, which shows a weak dependence of critical temperature on the chemical potential. We then calculate the temperature dependent Polyakov loop up to the chemical potential $\mu=0.8 \rm{GeV}$, and obtain the $T - \mu$ phase diagram for deconfinement phase transition as shown in Fig.\ref{Tmu1}. From this figure, one can see that the deconfinement transition temperature is always a crossover and its critical temperature depends weakly on the chemical potential. This finding is consistent with that in the PNJL model \cite{Li:2018ygx}.

\section{Chiral phase transition and the quarkyonic phase}
\label{chiral}

In previous sections, we have fixed the chemical potential dependence of the dilaton background from the equation of state and baryon number susceptibilities,  and we have investigated the deconfinement phase transition in the $(T,\mu)$ plane. In this section, we will discuss the chiral phase transition at finite temperature and chemical potential.

\subsection{Building $V(\chi, F^2)$}
 We will consider the action part in Eq.(\ref{SMatter}), which includes the dynamics of the scalar field $X$, corresponding to the chiral condensate $\sigma\equiv\langle\bar{q}q\rangle$. When the scalar field $X$ obtains a vacuum expectation value $X_0$, the $SU(2)_L \times SU(2)_R$ symmetry of the matter sector $S_M$ would be broken. We consider the two-flavor case $N_f = 2$ with $m_ u = m_d$, and we set $X_0(z) = \chi(z)I_2/2$ with $I_2$ the $2 \times 2$ identity matrix. From Eq.(\ref{SMatter}), the  action reduces to the form of
\begin{eqnarray}
S_\chi = - \int d^5x \sqrt{-g^s}e^{-\Phi}[\frac{1}{2}g^{zz}\chi^{\prime2}+V(\chi,F^2)],
\end{eqnarray}
where $V(\chi,F^2) \equiv {\rm Tr}(V_X(X,F^2))$. The equation of motion can be easily derived as
\begin{eqnarray}
\chi^{\prime\prime}(z) + (-\frac{3}{z} + 3A^\prime_s(z) - \phi^\prime(z) + \frac{f^\prime(z)}{f(z)})\chi^\prime(z)+ \frac{e^{2A_s(z)}}{z^2f(z)}\partial_\chi V(\chi(z),F^2) = 0.\label{eq-chi-1}
\end{eqnarray}

Generally, the potential $V_X$ could be of any form satisfying the $SU(2)_L\times SU(2)_R$ symmetry.   In previous studies \cite{Chelabi:2015cwn,Chelabi:2015gpc}, it is taken as a simple form $V_X(X,F^2)=M_5^2 X^+ X +\alpha (X^+X)^2$ with $M_5^2=-3$ from the AdS/CFT prescription, or equivalently
\begin{eqnarray}
V(\chi, F^2)=-\frac{3}{2}\chi^2+v_4 \chi^4.
\end{eqnarray}
It is shown that $v_4$ should be positive in order to have chiral symmetry breaking at low temperature in chiral limit. Therefore, firstly, we would have a test on this simple form of potential. Inserting this potential into Eq.(\ref{eq-chi-1}), one can obtain the leading UV($z\rightarrow 0$) expansion of $\chi(z)$ as
\begin{eqnarray}
\chi(z) = m_q\zeta z +...+ \frac{\sigma}{\zeta}z^3 + ...,
\end{eqnarray}
where $\sigma$ is the chiral condensate(order parameter) and $\zeta = \frac{\sqrt{3}}{2\pi}$ is the normalization constant obtain by matching correlation function$\langle\bar{q}q(p)\bar{q}q(0)\rangle$ with 4D calculation(for details, please refer to  \cite{Cherman:2008eh}). Near the horizon, the regular condition of $\chi$ requires $\frac{1}{f(z)}(f^\prime\chi^\prime + e^2A_s\partial_\chi V(\chi)/z^2)$ to be finite at $z=z_h$. Taking the metric solved in previous sections and imposing the above UV and IR boundary conditions, one can solve $\sigma$ with respect to $m_q$, $T$ and $\mu$.

\begin{figure}
\begin{center}
\epsfxsize=6.5 cm \epsfysize=4.5 cm \epsfbox{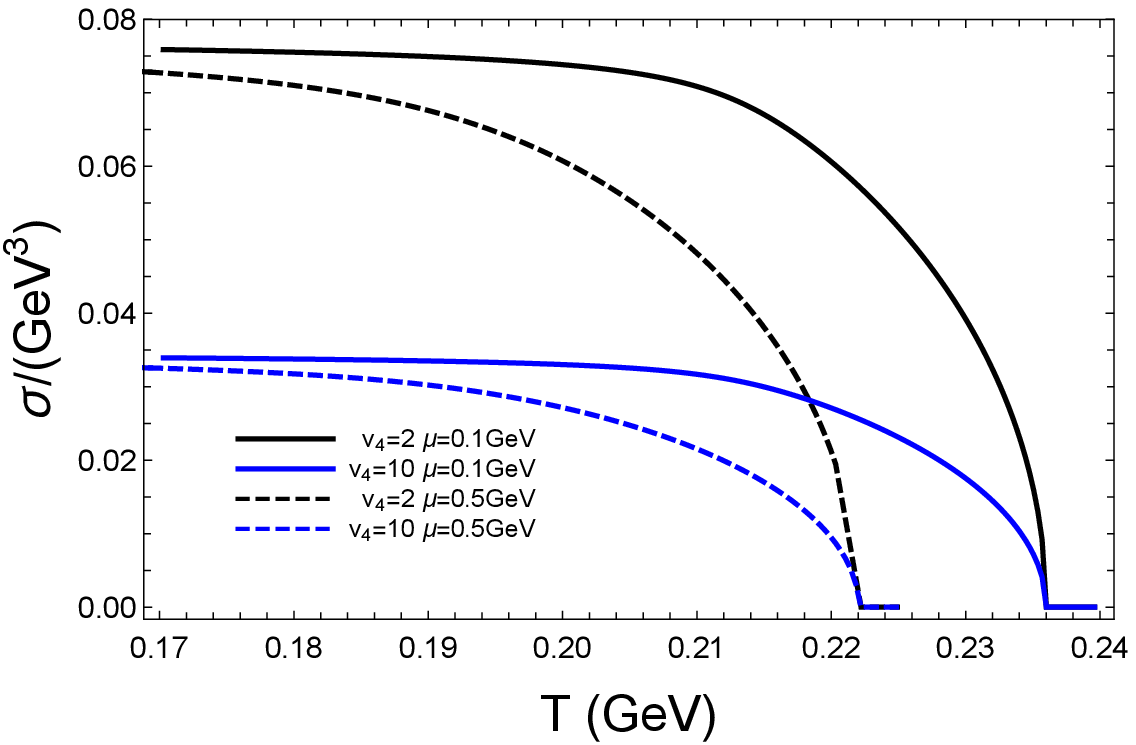}
\hspace*{0.1cm} \epsfxsize=6.5 cm \epsfysize=4.5 cm
\epsfbox{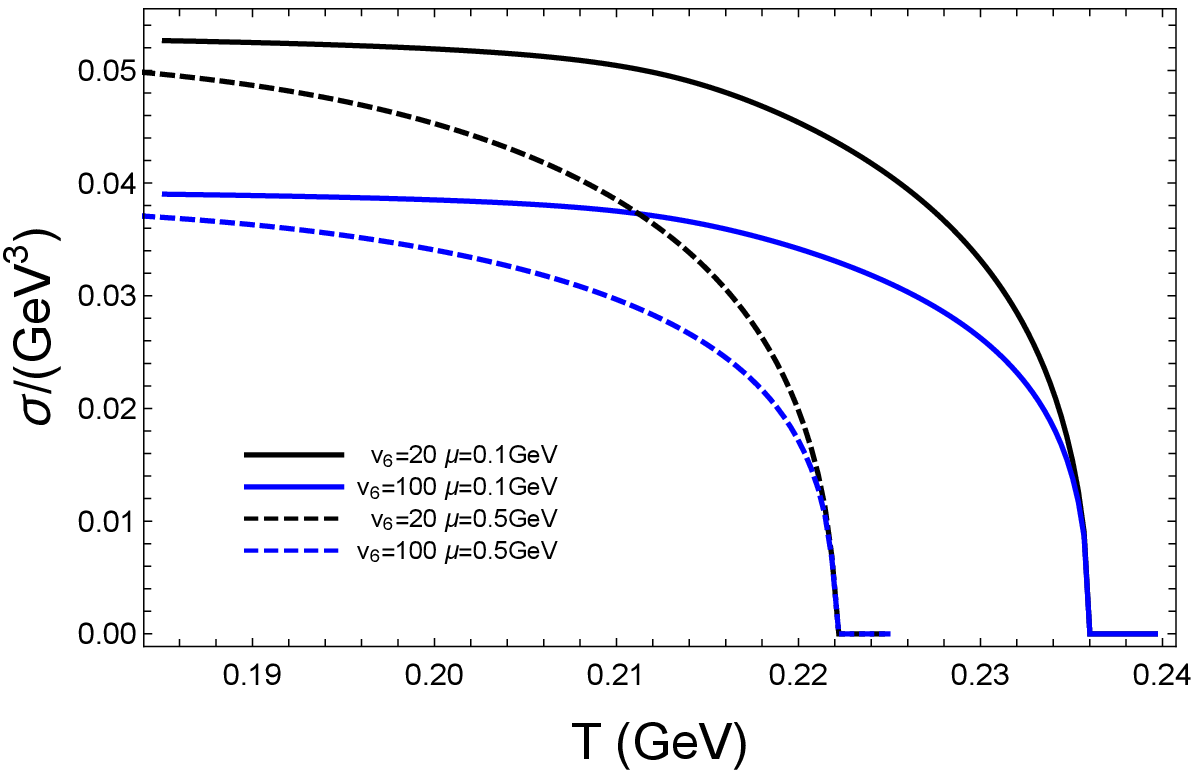} \vskip -0.05cm \hskip 0.15 cm
\textbf{( a ) } \hskip 6.5 cm \textbf{( b )} \\
\end{center}
\caption{Chiral condensate $\sigma$ in chiral limit($m_q=0$) as a function of the temperature T for different potentials $V(\chi, F^2)$ and chemical potentials. Panel.(a) gives the results for $V(\chi, F^2)=-\frac{3}{2}\chi^2+v_4\chi^4$ when $v_4=2,\mu=0.1\rm{GeV}$(black solid line), $v_4=2,\mu=0.5\rm{GeV}$( black dashed line), $v_4=10, \mu=0.1\rm{GeV}$(blue solid line), $v_4=10, \mu=0.5\rm{GeV}$(blue dashed line). Panel.(b) gives the results for $V(\chi, F^2)=-\frac{3}{2}\chi^2+v_6\chi^6$ when $v_6=20,\mu=0.1\rm{GeV}$(black solid line), $v_6=20,\mu=0.5\rm{GeV}$( black dashed line), $v_6=100, \mu=0.1\rm{GeV}$(blue solid line), $v_6=100, \mu=0.5\rm{GeV}$(blue dashed line). }
 \label{simplev4v6}
\end{figure}

Firstly we check whether the spontaneous symmetry breaking is described well. Therefore, we take $m_q=0$, i.e. in chiral limit. Then, we take different values of $v_4$ and $\mu$ and solve out $\sigma$ as a function of temperature in Fig.\ref{simplev4v6}(a)($v_4=2,\mu=0.1\rm{GeV}$(black solid line), $v_4=2,\mu=0.5\rm{GeV}$( black dashed line), $v_4=10, \mu=0.1\rm{GeV}$(blue solid line), $v_4=10, \mu=0.5\rm{GeV}$(blue dashed line)). From the figure, one could see that, for different values of $v_4$, the qualitative results for $\sigma$ are the same. At low temperature, chiral symmetry is broken by finite condensate $\sigma$, while above a certain temperature $T_c$, $\sigma$ turns to zero and chiral symmetry is restored. The phase transition is of second order type. It could also be seen that $T_c$ is not depend on $v_4$ for a fixed $\mu$\footnote{Actually, this could be easily understood by expanding Eq.(\ref{eq-chi-1}) to the leading order near $T_c$.(For details, see \cite{Chen:2018msc}).}.  Also, we find that a larger $v_4$ leads to a smaller $\sigma$ at the same temperature when $v_4>0$, while there is no condensate when $v_4\leq0$.

\begin{figure}
\centering
\includegraphics[width=8.5cm,height=6.5cm]{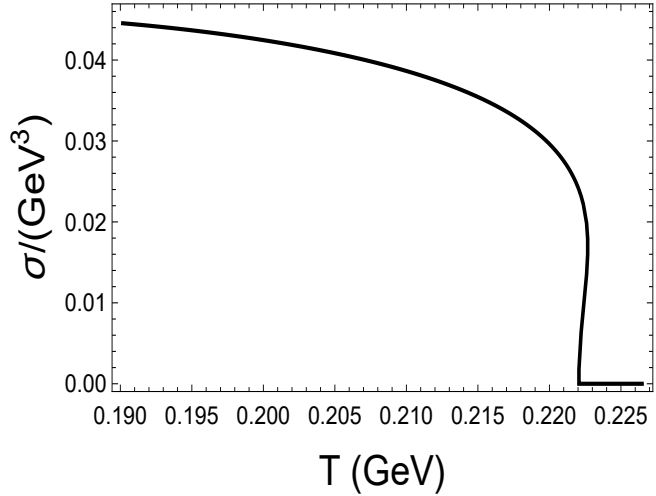}
\caption{Chiral condensate $\sigma$ in chiral limit($m_q=0$) as a function of the temperature T for  $V(\chi, F^2)=-\frac{3}{2}\chi^2+v_4\chi^4+v_6 \chi^6$ with $v_4=-2, v_6=100$ and $\mu=0.5\rm{GeV}$.}
\label{simplev4+v6}
\end{figure}

Up to finite temperature, it seems chiral phase transition could be well described by such a simple model. However, when one analyze the effect of baryon chemical potential $\mu$ one finds that the phase transition is always second order for any value of $\mu$. There is no first order transition region for any value of $\mu$. It is inconsistent with the expected phase diagram in $T-\mu$ plane for real QCD. Therefore, additional term in $V_X$ might be necessary. A direct extension is including the higher power $(X^+X)^3$ which satisfies the symmetry and considering effective potential
\begin{eqnarray}
V(\chi, F^2)=-\frac{3}{2}\chi^2+v_4 \chi^4+v_6\chi^6.
\end{eqnarray}

Similarly, we take chiral limit  to study the phase structure first. We take $v_4=0$ and different values of $v_6=20,\mu=0.1\rm{GeV}$(black solid line), $v_6=20,\mu=0.5\rm{GeV}$(black dashed line), $v_6=100, \mu=0.1\rm{GeV}$(blue solid line), $v_6=100, \mu=0.5\rm{GeV}$(blue dashed line) in Fig.\ref{simplev4v6}(b). From the figure, we find that the effect of $v_6$ is very similar to $v_4$. It will not affect the location of $T_c$, while it suppresses chiral condensate for a given value of $T$ and $\mu$. The transition order of chiral transition is still kept as second order one for any value of $\mu$.

Then we turn on the effect of $v_4, v_6$ simultaneously, and we find that when both $v_4$ and $v_6$ are positive, the qualitative results are almost the same as the above two cases. But if $v_4$ become negative, the solution structure near the transition point where $\sigma$ turns to zero becomes different. As an example, we take $v_4=-2, v_6=100, \mu=0.5\rm{GeV}$ and show the results in Fig.\ref{simplev4+v6}.  In this situation, we could see that a triple-valued region of $\sigma$ appears in a narrow temperature region $0.222\rm{GeV}<T<0.223\rm{GeV}$, which indicates a first order phase transition. Though the transition order depends only on the sign of $v_4$ and is independent on $\mu$, we see the possibility of constructing a model to describe different transition orders in different region of $\mu$ by changing the sign of the sub-leading power term $v_4$. If $v_4$ effectively changes its value from positive at small $\mu$ to negative at large $\mu$, one might expect a change of transition order. It is easy to see that if one adds a term like
\begin{eqnarray}
\alpha F^2(X^+X)^2
\end{eqnarray} to the action, effectively, the coefficient of $\chi^4$ term becomes \begin{eqnarray}
-\lambda_4A_t^{\prime2}\frac{z^4}{e^{4A_s}}+ v_4.
\end{eqnarray} Here, we have inserted $A_t=A_t(z), A_z=0$ and $A_i=0$ for spatial components of the gauge potential. When $\mu=0$, we take $v_4>0$, the transition is second order; while when $\mu$ is sufficient large, $-\lambda_4A_t^{\prime2}\frac{z^4}{e^{4A_s}}+ v_4$ will effectively become negative dominant at certain region and the transition order might turn to first order. Of course, generally, we can consider any power obeying the symmetry. Here we only consider the first powers $F^2(X^+X), F^2(X^+X)^2, F^2(X^+X)^3$ and assume the effect of higher power term would be suppressed by certain mechanism.

According to the above discussion, we will consider the following general form of potential $V(\chi,F^2)$

\begin{eqnarray}\label{vcf}
V(\chi,F^2) = (\frac{3}{2}  + \lambda_2A_t^{\prime2}\frac{z^4}{e^{4A_s}})\chi^2 \notag+ ( \lambda_4A_t^{\prime2}\frac{z^4}{e^{4A_s}}- v_4)\chi^4 + (\lambda_6A_t^{\prime2}\frac{z^4}{e^{4A_s}} - v_6)\chi^6.
\end{eqnarray}
Here, $v_4, v_6$ terms are corresponding to $(X^+X)^2, (X^+X)^3$ term, $\lambda_2, \lambda_4,\lambda_6$ terms are corresponding to $F^2 X^+X, F^2(X^+X)^2, F^2(X^+X)^3$ terms. We have imposed that at finite $\mu$ only $A_t(z)$ is non-vanishing and replaced $F^2$ with it. 

\begin{figure}
\begin{center}
\epsfxsize=6.5 cm \epsfysize=4.5 cm \epsfbox{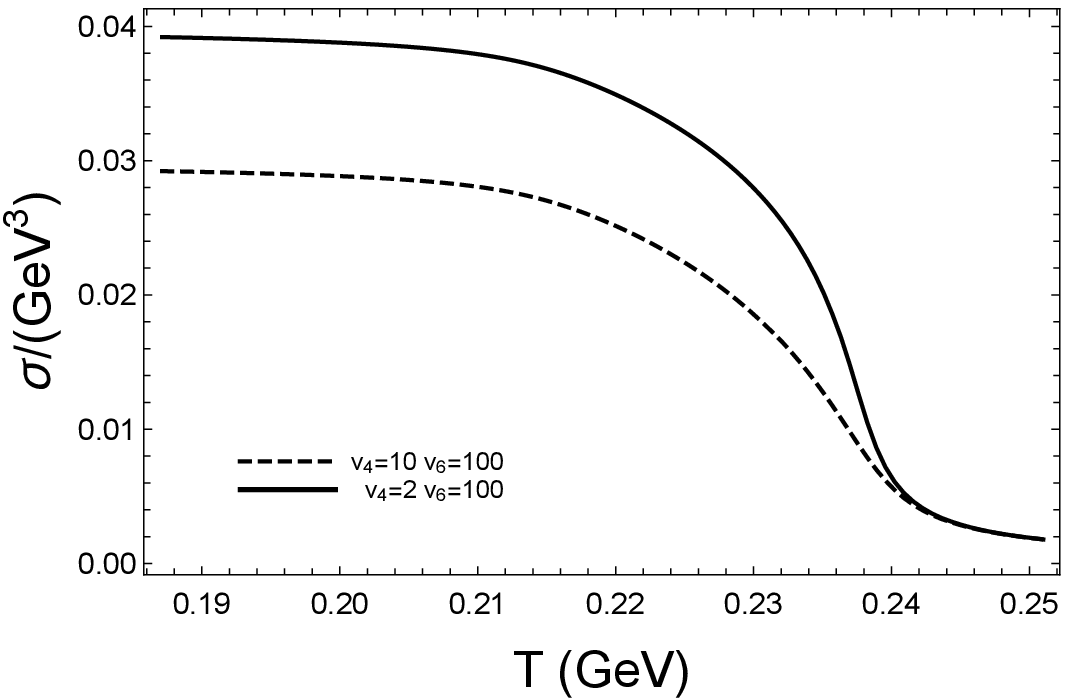}
\hspace*{0.1cm} \epsfxsize=6.5 cm \epsfysize=4.5 cm
\epsfbox{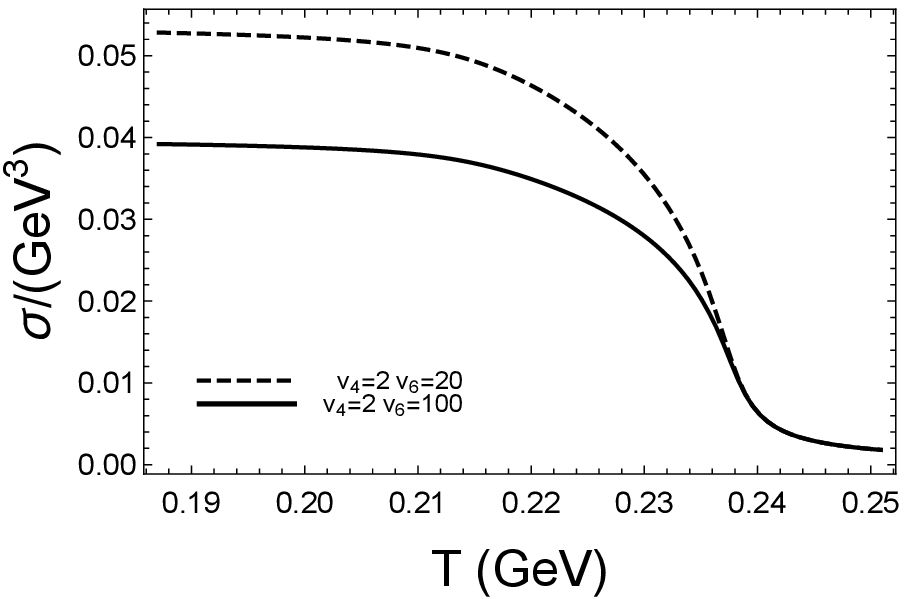} \vskip -0.05cm \hskip 0.15 cm
\textbf{( a ) } \hskip 6.5 cm \textbf{( b )} \\
\end{center}
\caption{Chiral condensate $\sigma$ as a function of the temperature T for $\mu=0$ and $m_q=5\rm{MeV}$, with $V(\chi, F^2)$ taking the form of Eq.(\ref{vcf}). Panel(a) gives results for $v_4=2, v_6=100$(solid line) and $v_4=10, v_6=100$(dashed line). Panel(b gives ) gives results for $v_4=2, v_6=100$(solid line) and $v_4=2, v_6=20$(dashed line). Here, since $\mu=0$, the values of $\lambda_2, \lambda_4, \lambda_6$ will not affect $\sigma$. }
 \label{v4-v6-mu=0}
\end{figure}

Therefore, up to now, from the analysis in chiral limit, we have introduced five parameters $v_4, v_6$ and $\lambda_2, \lambda_4, \lambda_6$ to describe chiral phase transition. Now we can start to describe the phenomenology of real QCD. So in the rest part we will take $m_q=0.005\rm{GeV}$, close to the physical value of $u,d$ quarks. Note that when $\mu=0$, $A_t\equiv0$ and $\lambda_2,\lambda_4,\lambda_6$ will not enter the equation of motion of $\chi$. In this situation, only $v_4, v_6$ affect the behavior of $\sigma$. In Fig.\ref{v4-v6-mu=0}, we show the effects of $v_4, v_6$ on $\sigma$ when $\mu=0$. From the figure, we could see that, since $m_q\neq0$, the phase transition turns from second order to crossover type. Different from chiral limit, $\sigma$ approaches to zero at high temperature instead of being exactly zero at $T_c$. Here, we simply define the pseudo transition temperature $T_c$ as $\frac{\partial^2 \sigma}{\partial^2 T}|_{T_c}=0$. From the figure, we can see that the larger value of $v_4$ or $v_6$ leads to a smaller value of $\sigma$. Also, they would change the pseudo transition temperature. Thus, changing $v_4, v_6$ we can get different values of vacuum condensate $\sigma_0\equiv \sigma(T=0,\mu=0) $ and $T_c$.  Therefore, by fitting $\sigma=(0.340\rm{GeV})^3, T_c=0.238GeV$,  close to the results with vanishing $\mu$ from previous studies\cite{Burger2013The,McLerran:2008ua}, we can fix $v_4=2, v_6=100$ firstly. Then, only three free parameters $\lambda_2, \lambda_4, \lambda_6$ are left.

\begin{figure}
\begin{center}
\epsfxsize=6.5 cm \epsfysize=4.5 cm \epsfbox{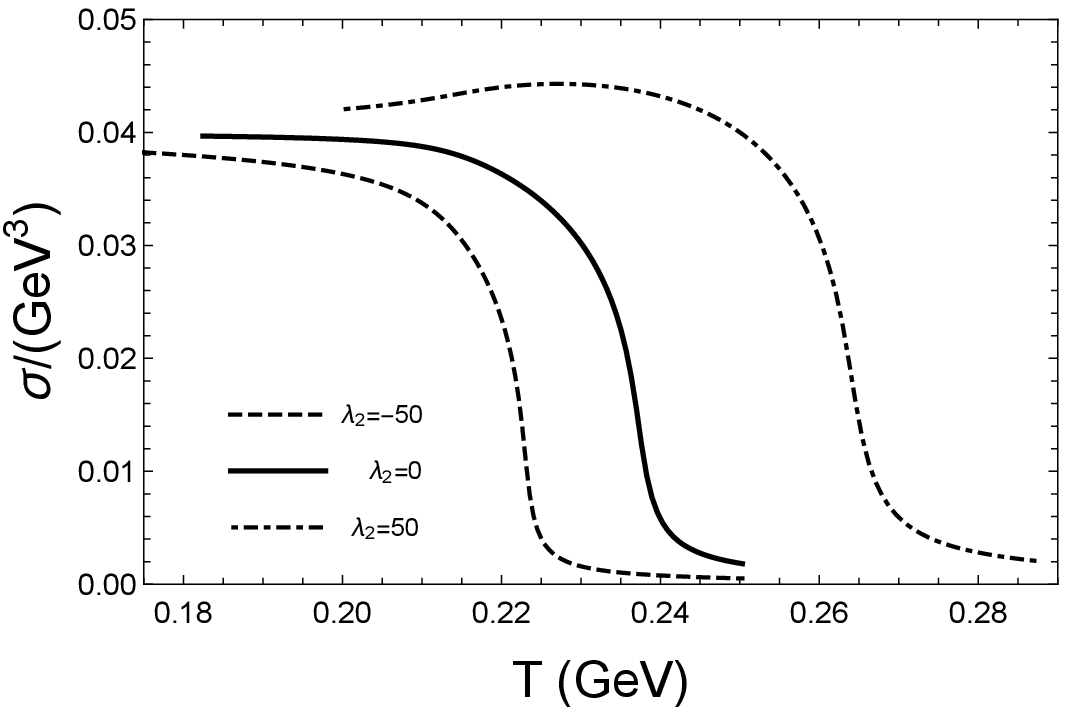}
\hspace*{0.1cm} \epsfxsize=6.5 cm \epsfysize=4.5 cm \epsfbox{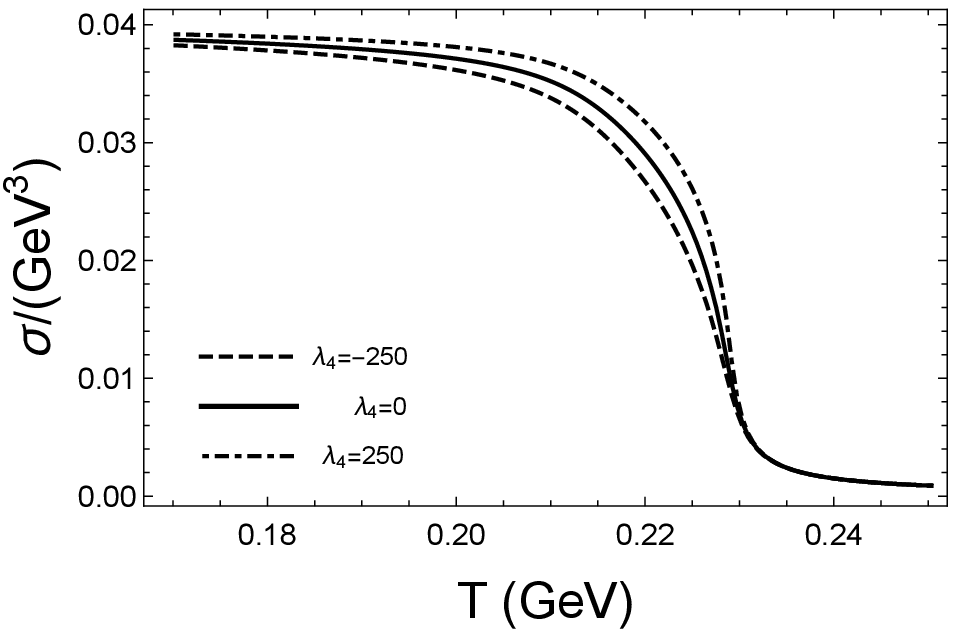}
\hspace*{0.1cm} \epsfxsize=6.5 cm \epsfysize=4.5 cm
\epsfbox{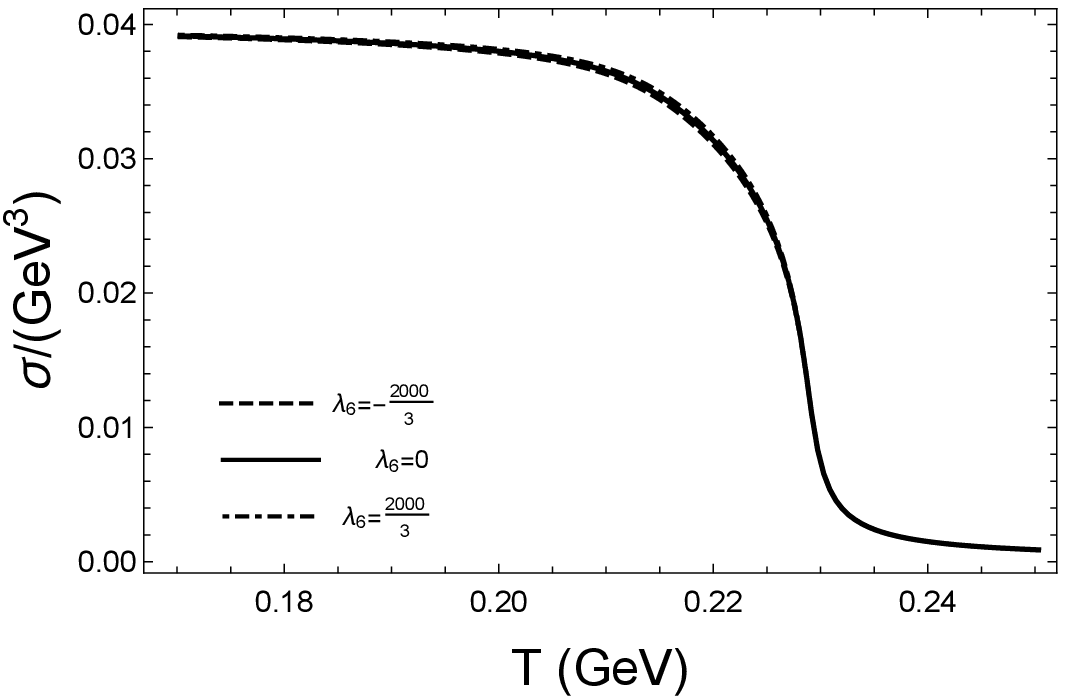} \vskip -0.05cm \hskip 0.15 cm
 \\
\end{center}
\caption{The chiral condensate $\sigma$ as a function of T for different parameters at $\mu = 0.1GeV$. (a) The solid line is for $\lambda_2 = 0$, the dashed line is for $\lambda_2 = -50$ and the dot-dashed line is for $\lambda_2 = 50$ for fixed $\lambda_4 = 200$ and $\lambda_6 = -\frac{1000}{3}$, respectively. (b) The solid line is for $\lambda_4 = 0$, the dashed line is for $\lambda_4 = -250$ and the dot-dashed line is for $\lambda_4 = 250$ for fixed $\lambda_2 = -25$ and $\lambda_6 = -\frac{1000}{3}$, respectively. (c) The solid line is for $\lambda_6 = 0$, the dashed line is for $\lambda_6 = -\frac{2000}{3}$ and the dot-dashed line is for $\lambda_6 = \frac{2000}{3}$ for fixed $\lambda_2 = -25$ and $\lambda_4 = 200$, respectively.
 }
 \label{lambdas}
\end{figure}

Generally, if one works in the full back-reaction scenario, the parameters $\lambda_2, \lambda_4, \lambda_6$ will back-reacting the background geometric and also the equation of states. It could affect the baryon number susceptibilities $\chi_n^B$ even at $\mu=0$, since $\chi_n^B$ depends on the derivative with respect to $\mu$. In this way, one might fix $\lambda_2,\lambda_4,\lambda_6$ by fitting the lattice results at $\mu=0$. However, in this work, we neglect the back-reaction. Therefore, we can not fix these three parameters in this way. In Fig.\ref{lambdas}, we show the effects of $\lambda$s on chiral condensate in wide parameter regions, $-50\leq \lambda_2 \leq50$, $-250\leq \lambda_4 \leq250$, $-\frac{-2000}{3}\leq \lambda_6 \leq\frac{2000}{3}$. From Fig.\ref{lambdas}, we could see that the parameters $\lambda$ is the coupling strength of chiral condensate with chemical potential and the sensitivity of $\lambda$ is presented. It is found that the effect of $\lambda_2$ is dominated, comparing with high order coupling term $\lambda_4$ and $\lambda_6$. The positive/negative sign in front of $\lambda$ will increase/decrease the transition temperature at the same chemical potential, which also means the location of CEP will appear at lower/higher chemical potential in $T - \mu$ plane. Without the coupling of chiral condensate with chemical potential, the deconfinement transition and chiral phase transition line will almost overlap. Thus, we choose these certain set of $\lambda$ to show the qualitative behavior of coupling strength in this model. Therefore, in the later calculation we will take a specific group of parameters $\lambda_2 =-25$, $\lambda_4 =200$, $\lambda_6 =-\frac{1000}{3}$ to show the qualitative results and the approximate quantitative results to get a close CEP position of PNJL model. But we emphasize that the exact value of this parameters could be fixed in a full back-reaction scenario and the position of CEP could be predicted in that model, which will be left for the future study.

\subsection{Interplay of chiral and deconfinement phase transition: quarkyonic phase}
\begin{figure}
\centering
    \includegraphics[width=6.5cm,height=4.5cm]{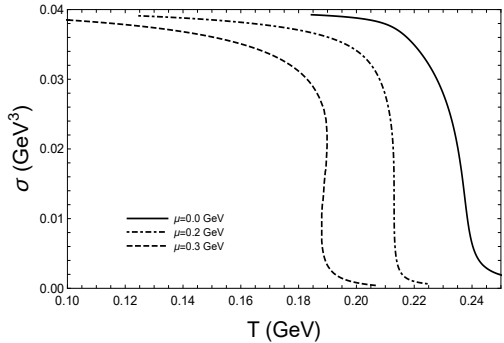}
\caption{The chiral condensate $\sigma$ as a function of the temperature T for three different quark chemical potentials $\mu$. The solid line is for $\mu = 0\ {\rm GeV}$, the dashed-dotted line is for $\mu = 0.2\ {\rm GeV}$ and the dashed line is for $\mu = 0.3\ {\rm GeV}$, respectively. The crossover becomes first order transition at high chemical potential.
 }
 \label{sigma}
\end{figure}

\begin{figure}
\centering
    \includegraphics[width=6.5cm,height=4.5cm]{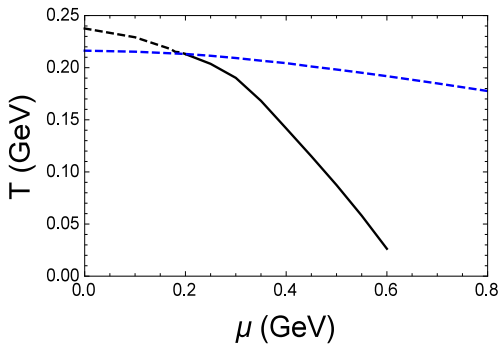}
\caption{The $(T,\mu)$ phase diagram for chiral and deconfinement phase transitions. The blue dashed line is for deconfinement phase transition and the black line is for the chiral phase transition, respectively.
 }
 \label{Tmu2}
\end{figure}

According to the above discussion, finally, we will take the parameters $\lambda_0=0, \lambda_2 =-25$, $\lambda_4 =200$, $\lambda_6 =-\frac{1000}{3}, v_4 =2$, $v_6 =100$ in our calculation.  Taking the physical quark mass $m_q=5\rm{MeV}$, we could solve out chiral condensate $\sigma$ as function of $T$ and $\mu$. We find that at small chemical potential, the phase transition is of crossover type, like the dashed line  for $\mu=0$ in Fig.\ref{sigma}. At large chemical potential, it is found that the phase transition turns to first order one, with multi-valued region appears, like the solid line for $\mu=0.3\rm{GeV}$ in Fig.\ref{sigma}.  In between the two cases, there is a critical point, where the phase transition turns to second order type, like the dotdashed line with $\mu^E=0.21\rm{GeV}$ in Fig.\ref{sigma}. One could define the second order transition temperature at $\frac{\partial\sigma}{\partial T}=\infty$, and we get $T^E=0.20\rm{GeV}$. So, we get when $\mu<\mu^E$, the phase transition is crossover type while when $\mu>\mu^E$ it turns to be first order one. In between them, it is the critical end point $(T^E,\mu^E)=(0.20,0.21)\rm{GeV}$ or $(T^E,\mu_B^E)=(0.20,0.63)\rm{GeV}$.
One could see that the critical baryon chemical potential at CEP in the quenched dynamical holographic QCD model is in good agreement with that in the realistic PNJL model \cite{Li:2018ygx}. Finally, varying chemical potential and extracting the corresponding critical temperature, one could draw the phase transition line for chiral phase transition in the black lines in Fig.\ref{Tmu2}.  

We summarize the chiral and deconfinement phase transitions in the $(T,\mu)$ plane in Fig.\ref{Tmu2}. Both the deconfinement and chiral phase transitions are realized in the quenched dynamical holographic QCD model. The deconfinement phase transition is always a crossover and it shows weak dependence on the quark chemical potential, and the chiral phase transition is a crossover at low chemical potential and turns to a first order phase transition at high chemical potential, and a CEP shows up at $(T^E,\mu^E)=(0.20,0.21)\rm{GeV}$ on the chiral phase transition line. The chiral phase transition has much stronger dependence on the quark chemical potential than the deconfinement phase transition, thus one can observe the quarkyonic phase showing up in the region of large chemical potential. This phase diagram is in agreement with that in the PNJL model \cite{McLerran:2008ua,Abuki:2008nm,Li:2018ygx}.

\section{Conclusion}
\label{sec-sum}

In this work, we investigate both the chiral and deconfinement phase transitions at finite temperature and chemical potential in the quenched dynamical holographic QCD model. In this quenched dynamical holographic QCD model, the dilaton background describes the gluodynamics and the flavor/meson background describes the chiral dynamics, respectively. To extend the quenched dynamical holographic QCD model to finite chemical potential, the quark chemical potential is introduced by a $U(1)$ field in the Einstein-Dilaton-Maxwell framework. The chemical potential dependence of the dilaton/gluodynamics is fixed by higher order baryon number fluctuations especially the kurtosis of baryon number fluctuations.  For the matter sector, we introduce a sextic term in the scalar potential to realize the first order phase transition at high chemical potential.

The chiral and deconfinement phase transitions in the $(T,\mu)$ plane is qualitatively consistent with that in the PNJL model. The deconfinement phase transition is always a crossover in the $(T,\mu)$ plane and it shows weak dependence on the quark chemical potential. The chiral phase transition is a crossover at low chemical potential and turns to be a first order phase transition at high chemical potential with a CEP showing up at $(T^E,\mu^E)=(0.20,0.21)\rm{GeV}$. The chiral phase transition has much stronger dependence on the quark chemical potential, therefore a quarkyonic phase with chiral symmetry restoration but still in confinement showing up in the region of large chemical potential. It is not surprising that the quenched dynamical holographic QCD model shows qualitatively consistent result with the PNJL model, because the PNJL model is also a model of quenched gluon background plus quark dynamics.

Naturally, we will consider the back reaction of the matter part action on the quenched gluodynamic background. Due to the complexity of numerical calculations of the coupling of two actions, we will try to solve the phase structure of the full dynamical holographic QCD model in the future.

\vskip 0.5cm
{\bf Acknowledgement}
\vskip 0.2cm
D.L. is supported by the NSFC under Grant Nos. 11805084 and 11647141, D.F.H is supported by the Ministry of Science and Technology
of China (MSTC) under the ¡±973¡± Project No.2015CB856904(4) and by the NSFC under Grant No. 11735007, and M.H. is supported in part by the NSFC under Grant Nos. 11725523, 11735007, 11261130311 (CRC 110 by DFG and NSFC), Chinese Academy of Sciences under Grant No. XDPB09, and the start-up funding from University of Chinese Academy of Sciences(UCAS).


\begin{thebibliography}{99}
\bibitem{Polyakov:1978vu}
  A.~M.~Polyakov,
  Phys.\ Lett.\  B {\bf 72}, 477 (1978).

\bibitem{'tHooft:1977hy}
  G.~'t Hooft,
  Nucl.\ Phys.\  B {\bf 138}, 1 (1978).

\bibitem{Casher:1979vw}
  A.~Casher,
  Phys.\ Lett.\  B {\bf 83}, 395 (1979).

\bibitem{Banks:1979yr}
  T.~Banks and A.~Casher,
  Nucl.\ Phys.\  B {\bf 169}, 103 (1980).

\bibitem{Hatta:2003ga}
  Y.~Hatta and K.~Fukushima,
  Phys.\ Rev.\  D {\bf 69}, 097502 (2004).

\bibitem{Mocsy:2003qw}
  A.~Mocsy, F.~Sannino and K.~Tuominen,
  Phys.\ Rev.\ Lett.\  {\bf 92}, 182302 (2004).

\bibitem{Braun:2007bx}
  J.~Braun, H.~Gies and J.~M.~Pawlowski,
  Phys.\ Lett.\  B {\bf 684}, 262 (2010).

\bibitem{McLerran:2007qj}
  L.~McLerran and R.~D.~Pisarski,
  Nucl.\ Phys.\ A {\bf 796}, 83 (2007)
  doi:10.1016/j.nuclphysa.2007.08.013
  [arXiv:0706.2191 [hep-ph]].

\bibitem{Hidaka:2008yy}
  Y.~Hidaka, L.~D.~McLerran and R.~D.~Pisarski,
  Nucl.\ Phys.\ A {\bf 808}, 117 (2008)
  doi:10.1016/j.nuclphysa.2008.05.009
  [arXiv:0803.0279 [hep-ph]].

\bibitem{McLerran:2008ua}
  L.~McLerran, K.~Redlich and C.~Sasaki,
  Nucl.\ Phys.\ A {\bf 824}, 86 (2009)
  doi:10.1016/j.nuclphysa.2009.04.001
  [arXiv:0812.3585 [hep-ph]].

\bibitem{Aoki:2006br}
  Y.~Aoki, Z.~Fodor, S.~D.~Katz and K.~K.~Szabo,
  Phys.\ Lett.\ B {\bf 643}, 46 (2006)
  doi:10.1016/j.physletb.2006.10.021
  [hep-lat/0609068].

\bibitem{Schmidt:2006us}
  C.~Schmidt,
  PoS LAT {\bf 2006}, 021 (2006)
  doi:10.22323/1.032.0021
  [hep-lat/0610116].

\bibitem{Philipsen:2005mj}
  O.~Philipsen,
  PoS LAT {\bf 2005}, 016 (2006)
  [PoS JHW {\bf 2005}, 012 (2006)]
  doi:10.22323/1.020.0016
  [hep-lat/0510077].

\bibitem{Heller:2006ub}
  U.~M.~Heller,
  PoS LAT {\bf 2006}, 011 (2006)
  doi:10.22323/1.032.0011
  [hep-lat/0610114].

\bibitem{Alkofer:2000wg}
  R.~Alkofer and L.~von Smekal,
  ``The Infrared behavior of QCD Green's functions: Confinement dynamical symmetry breaking, and hadrons as relativistic bound states,''
  Phys.\ Rept.\  {\bf 353}, 281 (2001)  [hep-ph/0007355].  


\bibitem{Bashir:2012fs}
  A.~Bashir, L.~Chang, I.~C.~Cloet, B.~El-Bennich, Y.~-X.~Liu, C.~D.~Roberts and P.~C.~Tandy,
  ``Collective perspective on advances in Dyson-Schwinger Equation QCD,''
  Commun.\ Theor.\ Phys.\  {\bf 58}, 79 (2012)  [arXiv:1201.3366 [nucl-th]].  


\bibitem{Wetterich:1992yh}
  C.~Wetterich,
  ``Exact evolution equation for the effective potential,''
  Phys.\ Lett.\ B {\bf 301}, 90 (1993).  

\bibitem{Pawlowski:2005xe}
  J.~M.~Pawlowski,
  ``Aspects of the functional renormalisation group,''
  Annals Phys.\  {\bf 322}, 2831 (2007)  [hep-th/0512261].  

\bibitem{Gies:2006wv}
  H.~Gies,
  ``Introduction to the functional RG and applications to gauge theories,''
  Lect.\ Notes Phys.\  {\bf 852}, 287 (2012)  [hep-ph/0611146].  


\bibitem{Nambu:1961fr}
  Y.~Nambu and G.~Jona-Lasinio,
  Phys.\ Rev.\  {\bf 124}, 246 (1961).
  doi:10.1103/PhysRev.124.246

\bibitem{Klevansky:1992qe}
  S.~P.~Klevansky,
  Rev.\ Mod.\ Phys.\  {\bf 64}, 649 (1992).
  doi:10.1103/RevModPhys.64.649

\bibitem{Meisinger:1995ih}
  P.~N.~Meisinger and M.~C.~Ogilvie,
  Phys.\ Lett.\ B {\bf 379}, 163 (1996)
  doi:10.1016/0370-2693(96)00447-9
  [hep-lat/9512011].

\bibitem{Fukushima:2003fw}
  K.~Fukushima,
  Phys.\ Lett.\ B {\bf 591}, 277 (2004)
  doi:10.1016/j.physletb.2004.04.027
  [hep-ph/0310121].

\bibitem{Ratti:2005jh}
  C.~Ratti, M.~A.~Thaler and W.~Weise,
  Phys.\ Rev.\ D {\bf 73}, 014019 (2006)
  doi:10.1103/PhysRevD.73.014019
  [hep-ph/0506234].

\bibitem{Roessner:2006xn}
  S.~Roessner, C.~Ratti and W.~Weise,
  Phys.\ Rev.\ D {\bf 75}, 034007 (2007)
  doi:10.1103/PhysRevD.75.034007
  [hep-ph/0609281].

\bibitem{Ghosh:2006qh}
  S.~K.~Ghosh, T.~K.~Mukherjee, M.~G.~Mustafa and R.~Ray,
  Phys.\ Rev.\ D {\bf 73}, 114007 (2006)
  doi:10.1103/PhysRevD.73.114007
  [hep-ph/0603050].

\bibitem{Schaefer:2007pw}
  B.~J.~Schaefer, J.~M.~Pawlowski and J.~Wambach,
  Phys.\ Rev.\ D {\bf 76}, 074023 (2007)
  doi:10.1103/PhysRevD.76.074023
  [arXiv:0704.3234 [hep-ph]].

\bibitem{Ratti:2007jf}
  C.~Ratti, S.~Roessner and W.~Weise,
  Phys.\ Lett.\ B {\bf 649}, 57 (2007)
  doi:10.1016/j.physletb.2007.03.038
  [hep-ph/0701091].

\bibitem{Sasaki:2006ww}
  C.~Sasaki, B.~Friman and K.~Redlich,
  Phys.\ Rev.\ D {\bf 75}, 074013 (2007)
  doi:10.1103/PhysRevD.75.074013
  [hep-ph/0611147].


\bibitem{Sasaki:2006ws}
  C.~Sasaki, B.~Friman and K.~Redlich,
  Phys.\ Rev.\ D {\bf 75}, 054026 (2007)
  doi:10.1103/PhysRevD.75.054026
  [hep-ph/0611143].

\bibitem{Megias:2006bn}
  E.~Megias, E.~Ruiz Arriola and L.~L.~Salcedo,
  Phys.\ Rev.\ D {\bf 74}, 114014 (2006)
  doi:10.1103/PhysRevD.74.114014
  [hep-ph/0607338].

\bibitem{Megias:2004hj}
  E.~Megias, E.~Ruiz Arriola and L.~L.~Salcedo,
  Phys.\ Rev.\ D {\bf 74}, 065005 (2006)
  doi:10.1103/PhysRevD.74.065005
  [hep-ph/0412308].

\bibitem{Zhang:2006gu}
  Z.~Zhang and Y.~X.~Liu,
  Phys.\ Rev.\ C {\bf 75}, 064910 (2007)
  doi:10.1103/PhysRevC.75.064910
  [hep-ph/0610221].

\bibitem{Sakai:2008py}
  Y.~Sakai, K.~Kashiwa, H.~Kouno and M.~Yahiro,
  Phys.\ Rev.\ D {\bf 77}, 051901 (2008)
  doi:10.1103/PhysRevD.77.051901
  [arXiv:0801.0034 [hep-ph]].


\bibitem{Ciminale:2007ei}
  M.~Ciminale, G.~Nardulli, M.~Ruggieri and R.~Gatto,
  Phys.\ Lett.\ B {\bf 657}, 64 (2007)
  doi:10.1016/j.physletb.2007.10.012
  [arXiv:0706.4215 [hep-ph]].


\bibitem{Fu:2007xc}
  W.~j.~Fu, Z.~Zhang and Y.~x.~Liu,
  Phys.\ Rev.\ D {\bf 77}, 014006 (2008)
  doi:10.1103/PhysRevD.77.014006
  [arXiv:0711.0154 [hep-ph]].

\bibitem{Hansen:2006ee}
  H.~Hansen, W.~M.~Alberico, A.~Beraudo, A.~Molinari, M.~Nardi and C.~Ratti,
  Phys.\ Rev.\ D {\bf 75}, 065004 (2007)
  doi:10.1103/PhysRevD.75.065004
  [hep-ph/0609116].

\bibitem{Contrera:2007wu}
  G.~A.~Contrera, D.~Gomez Dumm and N.~N.~Scoccola,
  Phys.\ Lett.\ B {\bf 661}, 113 (2008)
  doi:10.1016/j.physletb.2008.01.069
  [arXiv:0711.0139 [hep-ph]].

\bibitem{Maldacena:1997re}
  J.~M.~Maldacena,
  Int.\ J.\ Theor.\ Phys.\  {\bf 38}, 1113 (1999)
  [Adv.\ Theor.\ Math.\ Phys.\  {\bf 2}, 231 (1998)]
  doi:10.1023/A:1026654312961, 10.4310/ATMP.1998.v2.n2.a1
  [hep-th/9711200].

\bibitem{Gubser:1998bc}
  S.~S.~Gubser, I.~R.~Klebanov and A.~M.~Polyakov,
  Phys.\ Lett.\ B {\bf 428}, 105 (1998)
  doi:10.1016/S0370-2693(98)00377-3
  [hep-th/9802109].


\bibitem{Witten1998Anti}
  E.~Witten,
  Adv.\ Theor.\ Math.\ Phys.\  {\bf 2}, 253 (1998)
  doi:10.4310/ATMP.1998.v2.n2.a2
  [hep-th/9802150].


\bibitem{Erlich:2005qh}
  J.~Erlich, E.~Katz, D.~T.~Son and M.~A.~Stephanov,
  Phys.\ Rev.\ Lett.\  {\bf 95}, 261602 (2005)
  doi:10.1103/PhysRevLett.95.261602
  [hep-ph/0501128].



\bibitem{deTeramond:2005su}
  G.~F.~de Teramond and S.~J.~Brodsky,
  Phys.\ Rev.\ Lett.\  {\bf 94}, 201601 (2005)
  doi:10.1103/PhysRevLett.94.201601
  [hep-th/0501022].

\bibitem{DaRold:2005mxj}
  L.~Da Rold and A.~Pomarol,
  Nucl.\ Phys.\ B {\bf 721}, 79 (2005)
  doi:10.1016/j.nuclphysb.2005.05.009
  [hep-ph/0501218].


\bibitem{Babington:2003vm}
  J.~Babington, J.~Erdmenger, N.~J.~Evans, Z.~Guralnik and I.~Kirsch,
  Phys.\ Rev.\ D {\bf 69}, 066007 (2004)
  doi:10.1103/PhysRevD.69.066007
  [hep-th/0306018].

\bibitem{Kruczenski:2003uq}
  M.~Kruczenski, D.~Mateos, R.~C.~Myers and D.~J.~Winters,
  JHEP {\bf 0405}, 041 (2004)
  doi:10.1088/1126-6708/2004/05/041
  [hep-th/0311270].


\bibitem{Sakai:2004cn}
  T.~Sakai and S.~Sugimoto,
  Prog.\ Theor.\ Phys.\  {\bf 113}, 843 (2005)
  doi:10.1143/PTP.113.843
  [hep-th/0412141].


\bibitem{Sakai:2005yt}
  T.~Sakai and S.~Sugimoto,
  Prog.\ Theor.\ Phys.\  {\bf 114}, 1083 (2005)
  doi:10.1143/PTP.114.1083
  [hep-th/0507073].

\bibitem{Csaki:2006ji}
  C.~Csaki and M.~Reece,
  JHEP {\bf 0705}, 062 (2007)
  doi:10.1088/1126-6708/2007/05/062
  [hep-ph/0608266].

\bibitem{Huang:2007fv}
  S.~He, M.~Huang, Q.~S.~Yan and Y.~Yang,
  Eur.\ Phys.\ J.\ C {\bf 66}, 187 (2010)
  doi:10.1140/epjc/s10052-010-1239-0
  [arXiv:0710.0988 [hep-ph]].


\bibitem{Gherghetta:2009ac}
  T.~Gherghetta, J.~I.~Kapusta and T.~M.~Kelley,
  Phys.\ Rev.\ D {\bf 79}, 076003 (2009)
  doi:10.1103/PhysRevD.79.076003
  [arXiv:0902.1998 [hep-ph]].

\bibitem{Kelley:2010mu}
  T.~M.~Kelley, S.~P.~Bartz and J.~I.~Kapusta,
  Phys.\ Rev.\ D {\bf 83}, 016002 (2011)
  doi:10.1103/PhysRevD.83.016002
  [arXiv:1009.3009 [hep-ph]].


\bibitem{Sui:2009xe}
  Y.~Q.~Sui, Y.~L.~Wu, Z.~F.~Xie and Y.~B.~Yang,
  Phys.\ Rev.\ D {\bf 81}, 014024 (2010)
  doi:10.1103/PhysRevD.81.014024
  [arXiv:0909.3887 [hep-ph]].

\bibitem{Sui:2010ay}
  Y.~Q.~Sui, Y.~L.~Wu and Y.~B.~Yang,
  Phys.\ Rev.\ D {\bf 83}, 065030 (2011)
  doi:10.1103/PhysRevD.83.065030
  [arXiv:1012.3518 [hep-ph]].

\bibitem{Li:2012ay}
  D.~Li, M.~Huang and Q.~S.~Yan,
  Eur.\ Phys.\ J.\ C {\bf 73}, 2615 (2013)
  doi:10.1140/epjc/s10052-013-2615-3
  [arXiv:1206.2824 [hep-th]].

\bibitem{Li:2013oda}
  D.~Li and M.~Huang,
  JHEP {\bf 1311}, 088 (2013)
  doi:10.1007/JHEP11(2013)088
  [arXiv:1303.6929 [hep-ph]].

\bibitem{Chen:2015zhh}
  Y.~Chen and M.~Huang,
  Chin.\ Phys.\ C {\bf 40}, no. 12, 123101 (2016)
  doi:10.1088/1674-1137/40/12/123101
  [arXiv:1511.07018 [hep-ph]].

\bibitem{Shuryak:2004cy}
  E.~V.~Shuryak,
  Nucl.\ Phys.\ A {\bf 750}, 64 (2005)
  doi:10.1016/j.nuclphysa.2004.10.022
  [hep-ph/0405066].

\bibitem{Tannenbaum:2006ch}
  M.~J.~Tannenbaum,
  Rept.\ Prog.\ Phys.\  {\bf 69}, 2005 (2006)
  doi:10.1088/0034-4885/69/7/R01
  [nucl-ex/0603003].


\bibitem{Policastro:2001yc}
  G.~Policastro, D.~T.~Son and A.~O.~Starinets,
  Phys.\ Rev.\ Lett.\  {\bf 87}, 081601 (2001)
  doi:10.1103/PhysRevLett.87.081601
  [hep-th/0104066].

\bibitem{Cai:2009zv}
  R.~G.~Cai, Z.~Y.~Nie, N.~Ohta and Y.~W.~Sun,
  Phys.\ Rev.\ D {\bf 79}, 066004 (2009)
  doi:10.1103/PhysRevD.79.066004
  [arXiv:0901.1421 [hep-th]].

\bibitem{Cai:2008ph}
  R.~G.~Cai, Z.~Y.~Nie and Y.~W.~Sun,
  Phys.\ Rev.\ D {\bf 78}, 126007 (2008)
  doi:10.1103/PhysRevD.78.126007
  [arXiv:0811.1665 [hep-th]].

\bibitem{Sin:2004yx}
  S.~J.~Sin and I.~Zahed,
  Phys.\ Lett.\ B {\bf 608}, 265 (2005)
  doi:10.1016/j.physletb.2005.01.020
  [hep-th/0407215].

\bibitem{Shuryak:2005ia}
  E.~Shuryak, S.~J.~Sin and I.~Zahed,
  J.\ Korean Phys.\ Soc.\  {\bf 50}, 384 (2007)
  doi:10.3938/jkps.50.384
  [hep-th/0511199].

\bibitem{Janik:2005zt}
  R.~A.~Janik and R.~B.~Peschanski,
  Phys.\ Rev.\ D {\bf 73}, 045013 (2006)
  doi:10.1103/PhysRevD.73.045013
  [hep-th/0512162].

\bibitem{Nakamura:2006ih}
  S.~Nakamura and S.~J.~Sin,
  JHEP {\bf 0609}, 020 (2006)
  doi:10.1088/1126-6708/2006/09/020
  [hep-th/0607123].

\bibitem{Sin:2006pv}
  S.~J.~Sin, S.~Nakamura and S.~P.~Kim,
  JHEP {\bf 0612}, 075 (2006)
  doi:10.1088/1126-6708/2006/12/075
  [hep-th/0610113].

\bibitem{Herzog:2006gh}
  C.~P.~Herzog, A.~Karch, P.~Kovtun, C.~Kozcaz and L.~G.~Yaffe,
  JHEP {\bf 0607}, 013 (2006)
  doi:10.1088/1126-6708/2006/07/013
  [hep-th/0605158].

\bibitem{Gubser:2006bz}
  S.~S.~Gubser,
  Phys.\ Rev.\ D {\bf 74}, 126005 (2006)
  doi:10.1103/PhysRevD.74.126005
  [hep-th/0605182].

\bibitem{Zhang:2012jd}
  Z.~q.~Zhang, D.~f.~Hou and H.~c.~Ren,
  JHEP {\bf 1301}, 032 (2013)
  doi:10.1007/JHEP01(2013)032
  [arXiv:1210.5187 [hep-th]].

\bibitem{Li:2014hja}
  D.~Li, J.~Liao and M.~Huang,
  Phys.\ Rev.\ D {\bf 89}, no. 12, 126006 (2014)
  doi:10.1103/PhysRevD.89.126006
  [arXiv:1401.2035 [hep-ph]].

\bibitem{Li:2014dsa}
  D.~Li, S.~He and M.~Huang,
  JHEP {\bf 1506}, 046 (2015)
  doi:10.1007/JHEP06(2015)046
  [arXiv:1411.5332 [hep-ph]].

\bibitem{Aharony:1999ti}
  O.~Aharony, S.~S.~Gubser, J.~M.~Maldacena, H.~Ooguri and Y.~Oz,
  Phys.\ Rept.\  {\bf 323}, 183 (2000)
  doi:10.1016/S0370-1573(99)00083-6
  [hep-th/9905111].

\bibitem{Erdmenger:2007cm}
  J.~Erdmenger, N.~Evans, I.~Kirsch and E.~Threlfall,
  Eur.\ Phys.\ J.\ A {\bf 35}, 81 (2008)
  doi:10.1140/epja/i2007-10540-1
  [arXiv:0711.4467 [hep-th]].

\bibitem{Brodsky:2014yha}
  S.~J.~Brodsky, G.~F.~de Teramond, H.~G.~Dosch and J.~Erlich,
  Phys.\ Rept.\  {\bf 584}, 1 (2015)
  doi:10.1016/j.physrep.2015.05.001
  [arXiv:1407.8131 [hep-ph]].

\bibitem{Kim:2012ey}
  Y.~Kim, I.~J.~Shin and T.~Tsukioka,
  Prog.\ Part.\ Nucl.\ Phys.\  {\bf 68}, 55 (2013)
  doi:10.1016/j.ppnp.2012.09.002
  [arXiv:1205.4852 [hep-ph]].

\bibitem{Adams:2012th}
  A.~Adams, L.~D.~Carr, T.~Sch?fer, P.~Steinberg and J.~E.~Thomas,
  New J.\ Phys.\  {\bf 14}, 115009 (2012)
  doi:10.1088/1367-2630/14/11/115009
  [arXiv:1205.5180 [hep-th]].


\bibitem{Herzog:2006ra}
  C.~P.~Herzog,
  Phys.\ Rev.\ Lett.\  {\bf 98}, 091601 (2007)
  doi:10.1103/PhysRevLett.98.091601
  [hep-th/0608151].

\bibitem{BallonBayona:2007vp}
  C.~A.~Ballon Bayona, H.~Boschi-Filho, N.~R.~F.~Braga and L.~A.~Pando Zayas,
  Phys.\ Rev.\ D {\bf 77}, 046002 (2008)
  doi:10.1103/PhysRevD.77.046002
  [arXiv:0705.1529 [hep-th]].

\bibitem{Cai:2007zw}
  R.~G.~Cai and J.~P.~Shock,
  JHEP {\bf 0708}, 095 (2007)
  doi:10.1088/1126-6708/2007/08/095
  [arXiv:0705.3388 [hep-th]].


\bibitem{Cai:2012eh}
  R.~G.~Cai, S.~Chakrabortty, S.~He and L.~Li,
  JHEP {\bf 1302}, 068 (2013)
  doi:10.1007/JHEP02(2013)068
  [arXiv:1209.4512 [hep-th]].




\bibitem{Kim:2007em}
  Y.~Kim, B.~H.~Lee, S.~Nam, C.~Park and S.~J.~Sin,
  Phys.\ Rev.\ D {\bf 76}, 086003 (2007)
  doi:10.1103/PhysRevD.76.086003
  [arXiv:0706.2525 [hep-ph]].



\bibitem{Andreev:2009zk}
  O.~Andreev,
  Phys.\ Rev.\ Lett.\  {\bf 102}, 212001 (2009)
  doi:10.1103/PhysRevLett.102.212001
  [arXiv:0903.4375 [hep-ph]].

\bibitem{Colangelo:2010pe}
  P.~Colangelo, F.~Giannuzzi and S.~Nicotri,
  Phys.\ Rev.\ D {\bf 83}, 035015 (2011)
  doi:10.1103/PhysRevD.83.035015
  [arXiv:1008.3116 [hep-ph]].
\bibitem{Gubser:2008yx}
  S.~S.~Gubser, A.~Nellore, S.~S.~Pufu and F.~D.~Rocha,
  Phys.\ Rev.\ Lett.\  {\bf 101}, 131601 (2008)
  doi:10.1103/PhysRevLett.101.131601
  [arXiv:0804.1950 [hep-th]].

\bibitem{Gubser:2008ny}
  S.~S.~Gubser and A.~Nellore,
  Phys.\ Rev.\ D {\bf 78}, 086007 (2008)
  doi:10.1103/PhysRevD.78.086007
  [arXiv:0804.0434 [hep-th]].

\bibitem{Gubser:2008sz}
  S.~S.~Gubser, S.~S.~Pufu and F.~D.~Rocha,
  JHEP {\bf 0808}, 085 (2008)
  doi:10.1088/1126-6708/2008/08/085
  [arXiv:0806.0407 [hep-th]].

\bibitem{Gursoy:2008bu}
  U.~Gursoy, E.~Kiritsis, L.~Mazzanti and F.~Nitti,
  Phys.\ Rev.\ Lett.\  {\bf 101}, 181601 (2008)
  doi:10.1103/PhysRevLett.101.181601
  [arXiv:0804.0899 [hep-th]].

\bibitem{Gursoy:2007cb}
  U.~Gursoy and E.~Kiritsis,
  JHEP {\bf 0802}, 032 (2008)
  doi:10.1088/1126-6708/2008/02/032
  [arXiv:0707.1324 [hep-th]].

\bibitem{Gursoy:2007er}
  U.~Gursoy, E.~Kiritsis and F.~Nitti,
  JHEP {\bf 0802}, 019 (2008)
  doi:10.1088/1126-6708/2008/02/019
  [arXiv:0707.1349 [hep-th]].

\bibitem{Gursoy:2008za}
  U.~Gursoy, E.~Kiritsis, L.~Mazzanti and F.~Nitti,
  JHEP {\bf 0905}, 033 (2009)
  doi:10.1088/1126-6708/2009/05/033
  [arXiv:0812.0792 [hep-th]].


\bibitem{Yaresko:2013tia}
  R.~Yaresko and B.~Kampfer,
  Phys.\ Lett.\ B {\bf 747}, 36 (2015)
  doi:10.1016/j.physletb.2015.05.034
  [arXiv:1306.0214 [hep-ph]].

\bibitem{Li:2011hp}
  D.~Li, S.~He, M.~Huang and Q.~S.~Yan,
  JHEP {\bf 1109}, 041 (2011)
  doi:10.1007/JHEP09(2011)041
  [arXiv:1103.5389 [hep-th]].

\bibitem{Cai:2012xh}
  R.~G.~Cai, S.~He and D.~Li,
  JHEP {\bf 1203}, 033 (2012)
  doi:10.1007/JHEP03(2012)033
  [arXiv:1201.0820 [hep-th]].

\bibitem{He:2013qq}
  S.~He, S.~Y.~Wu, Y.~Yang and P.~H.~Yuan,
  JHEP {\bf 1304}, 093 (2013)
  doi:10.1007/JHEP04(2013)093
  [arXiv:1301.0385 [hep-th]].

\bibitem{Yang:2014bqa}
  Y.~Yang and P.~H.~Yuan,
  JHEP {\bf 1411}, 149 (2014)
  doi:10.1007/JHEP11(2014)149
  [arXiv:1406.1865 [hep-th]].

\bibitem{Zuo:2014iza}
  F.~Zuo,
  JHEP {\bf 1406}, 143 (2014)
  doi:10.1007/JHEP06(2014)143
  [arXiv:1404.4512 [hep-ph]].

\bibitem{Afonin:2014jha}
  S.~S.~Afonin and A.~D.~Katanaeva,
  Eur.\ Phys.\ J.\ C {\bf 74}, no. 10, 3124 (2014)
  doi:10.1140/epjc/s10052-014-3124-8
  [arXiv:1408.6935 [hep-ph]].

\bibitem{Bruckmann:2013oba}
  F.~Bruckmann, G.~Endrodi and T.~G.~Kovacs,
  JHEP {\bf 1304}, 112 (2013)
  doi:10.1007/JHEP04(2013)112
  [arXiv:1303.3972 [hep-lat]].

\bibitem{Rougemont:2015oea}
  R.~Rougemont, R.~Critelli and J.~Noronha,
  Phys.\ Rev.\ D {\bf 93}, no. 4, 045013 (2016)
  doi:10.1103/PhysRevD.93.045013
  [arXiv:1505.07894 [hep-th]].




\bibitem{Karch:2006pv}
  A.~Karch, E.~Katz, D.~T.~Son and M.~A.~Stephanov,
  Phys.\ Rev.\ D {\bf 74}, 015005 (2006)
  doi:10.1103/PhysRevD.74.015005
  [hep-ph/0602229].

\bibitem{Chelabi:2015cwn}
  K.~Chelabi, Z.~Fang, M.~Huang, D.~Li and Y.~L.~Wu,
  Phys.\ Rev.\ D {\bf 93}, no. 10, 101901 (2016)
  doi:10.1103/PhysRevD.93.101901
  [arXiv:1511.02721 [hep-ph]].

\bibitem{Chelabi:2015gpc}
  K.~Chelabi, Z.~Fang, M.~Huang, D.~Li and Y.~L.~Wu,
  JHEP {\bf 1604}, 036 (2016)
  doi:10.1007/JHEP04(2016)036
  [arXiv:1512.06493 [hep-ph]].

\bibitem{Fang:2015ytf}
  Z.~Fang, S.~He and D.~Li,
  Nucl.\ Phys.\ B {\bf 907}, 187 (2016)
  doi:10.1016/j.nuclphysb.2016.04.003
  [arXiv:1512.04062 [hep-ph]].

\bibitem{Evans:2016jzo}
  N.~Evans, C.~Miller and M.~Scott,
  Phys.\ Rev.\ D {\bf 94}, no. 7, 074034 (2016)
  doi:10.1103/PhysRevD.94.074034
  [arXiv:1604.06307 [hep-ph]].


\bibitem{Evans:2010iy}
  N.~Evans, A.~Gebauer, K.~Y.~Kim and M.~Magou,
  JHEP {\bf 1003}, 132 (2010)
  doi:10.1007/JHEP03(2010)132
  [arXiv:1002.1885 [hep-th]].

\bibitem{Evans:2012cx}
  N.~Evans, K.~Y.~Kim, M.~Magou, Y.~Seo and S.~J.~Sin,
  JHEP {\bf 1209}, 045 (2012)
  doi:10.1007/JHEP09(2012)045
  [arXiv:1204.5640 [hep-th]].

\bibitem{Evans:2011eu}
  N.~Evans, A.~Gebauer, M.~Magou and K.~Y.~Kim,
  J.\ Phys.\ G {\bf 39}, 054005 (2012)
  doi:10.1088/0954-3899/39/5/054005
  [arXiv:1109.2633 [hep-th]].

\bibitem{Sakai:2012ika}
  Y.~Sakai, H.~Kouno, T.~Sasaki and M.~Yahiro,
  Phys.\ Lett.\ B {\bf 718}, 130 (2012)
  doi:10.1016/j.physletb.2012.10.027
  [arXiv:1204.0228 [hep-ph]].



\bibitem{Li:2018ygx}
  Z.~Li, K.~Xu, X.~Wang and M.~Huang,
  Eur.\ Phys.\ J.\ C {\bf 79}, no. 3, 245 (2019)
  doi:10.1140/epjc/s10052-019-6703-x
  [arXiv:1801.09215 [hep-ph]].

\bibitem{Li:2017ple}
  Z.~Li, Y.~Chen, D.~Li and M.~Huang,
  Chin.\ Phys.\ C {\bf 42}, no. 1, 013103 (2018)
  doi:10.1088/1674-1137/42/1/013103
  [arXiv:1706.02238 [hep-ph]].

\bibitem{Bazavov2017The}
  Burger, F. and Ilgenfritz, E. M. and Kirchner, M. and Lombardo, M. P. and Mullerpreussker, M. and Philipsen, O. and Pinke, C. and Urbach, C. and Zeidlewicz, L.
  ``The QCD Equation of State to $\mathcal{O}(\mu_B^6)$ from Lattice QCD''
 ,  (2017) [arXiv:1701.04325v3 [hep-lat]]


\bibitem{Li:2013xpa}
  M.~Huang and D.~Li,
  Springer Proc.\ Phys.\  {\bf 170}, 367 (2016)
  [arXiv:1311.0593 [hep-ph]].

\bibitem{Li:2014txa}
  D.~Li and M.~Huang,
  EPJ Web Conf.\  {\bf 80}, 00011 (2014)
  doi:10.1051/epjconf/20148000011
  [arXiv:1409.8432 [hep-ph]].

\bibitem{tHooft:1993dmi}
  G.~'t Hooft,
  ``Dimensional reduction in quantum gravity,''
  Conf.\ Proc.\ C {\bf 930308} (1993) 284
  [gr-qc/9310026].

\bibitem{Susskind:1994vu}
  L.~Susskind,
  ``The World as a hologram,''
  J.\ Math.\ Phys.\  {\bf 36} (1995) 6377
  [hep-th/9409089].

\bibitem{Burger2015The}
 Burger, F. and Ilgenfritz, E. M. and Lombardo, M. P. and Mullerpreussker, M.
  ``The equation of state of quark-gluon matter from lattice QCD with two flavors of twisted-mass Wilson fermions''
  Phys.\ Rev.\  D  {\bf 91} (2015)


\bibitem{Burger2013The}
  Burger, F. and Ilgenfritz, E. M. and Kirchner, M. and Lombardo, M. P. and Mullerpreussker, M. and Philipsen, O. and Pinke, C. and Urbach, C. and Zeidlewicz, L.
  ``The thermal QCD transition with two flavours of twisted mass fermions''
 Phys.\ Rev.\  D {\bf 87},  (2013)

\bibitem{Cherman:2008eh}
  A.~Cherman, T.~D.~Cohen and E.~S.~Werbos,
  ``The Chiral condensate in holographic models of QCD,''  Phys.\ Rev.\ C {\bf 79} (2009) 045203  [arXiv:0804.1096 [hep-ph]].

\bibitem{Chen:2018msc}
  J.~Chen, S.~He, M.~Huang and D.~Li,
  ``Critical exponents of finite temperature chiral phase transition in soft-wall AdS/QCD models,''
  JHEP {\bf 1901} (2019) 165
  [arXiv:1810.07019 [hep-ph]].

\bibitem{Abuki:2008nm}
  H.~Abuki, R.~Anglani, R.~Gatto, G.~Nardulli and M.~Ruggieri,
  Phys.\ Rev.\ D {\bf 78}, 034034 (2008)
  doi:10.1103/PhysRevD.78.034034
  [arXiv:0805.1509 [hep-ph]].

\end{thebibliography}
\end{document}